\documentclass[preprint,aps,showpacs,showkeys,prb]{revtex4}

\usepackage{amssymb}
\usepackage{graphicx}
\usepackage{amsmath}
\usepackage{amsfonts}
\usepackage{latexsym}
\usepackage{natbib}
\usepackage{epsfig}

\begin{document}

\title{A generalized definition of spin in non-orientable geometries}

\author{A. Rebei}

\affiliation{Seagate Research Center\\
Pittsburgh, Pennsylvania 1522, USA\\ 
arebei@mailaps.org}

\email{arebei@mailaps.org}

\begin{abstract}
Non-orientable nanostructures are becoming feasable today. \ This 
lead us to the study of spin in these geometries. \ Hence a 
physically sound definition of spin is suggested. \ Using 
our definition, we study the question of the number of 
different ways to define spin. \ We argue that the possibility 
of having more than one spin structure should be taken into 
account energetically. \ The effect
of topology on spin is studied in detail using cohomological 
arguments. \ We generalize the definition of equivalence among (s)pin 
structures to include non-orientable spaces.
\end{abstract}


\maketitle

\section{INTRODUCTION}

 It has recently been possible to realize new small-size 
materials with nontrivial 
topologies. \ Tanda et al. were able to have
 Mobius bands formed by crystals of Niobium and Selenium, $NbSe_3$
\cite{japan}. \ Given the recent interest in spintronics,
 it seems therefore 
worth the effort to study the possible 
effect of geometry on 
spin, especially the combined effect of 
non-orientability and 
non-simple connectedness of the space.

\ Geometric effects in physics are often hidden 
in terms of constraints. \ One good example of this 
is the $\theta$ vacua of QCD. \ Here, the different 
topological sectors are due to the Gauss 
constraint \cite{ba}. \ The Chern number 
also appears due to constraints either 
in physical 
space or momentum space. \ While studying charge density
waves in a torus geometry, Thouless 
 found that in this case, 
the conductivity is expressed in terms of the Chern number
of the manifold and differs from the periodic lattice case 
in Euclidean space \cite{thouless}. \ Spin currents in 
semiconductors is still another problem where the 
Chern number can be used to explain the universality
of the spin conductivity in the Rashba model 
\cite{rebei}. \ In this latter case, the constraint is in momentum 
space which is homotopic to the plane without the origin, 
a non-simply connected space.  \ A final example, we give, is 
the quantization of the spin of the $SU(2)$ Skyrmion. \ The 
solution to this problem  was possible only 
after extending $SU(2)$ to 
group $SU(3)$ \cite{witten}. \ However in
this extension, a 
new 
term is needed in the Lagrangian, the Wess-Zumino 
term, which has a topological
significance and it is related to the disconnectedness 
of $\mathbf{SO}(3)$. \ This latter example shows the
inter-connectedness of topology of 
fields and spin.

\ In this work, we address similar issues 
between spin  and topology
of the physical space of 
 electrons on non-orientable manifolds such as a Mobius band. \ In 
this case 
there is 
no global well defined  spin structure for the 
manifold \cite{alvarez}. \ It is well
 known that  quantum mechanical wave functions
 in non-simply connected spaces can be multivalued and
are therefore  well
 defined only on their simply connected 
covering spaces \cite{ba}. \ For non-orientable manifolds, we 
show 
that a definition of spin is possible
 by going to the 
orientable double cover of the initial space. \ This work 
was motivated by Tanda's group and a simple calculation 
that is presented in the application section. \ For thin-film
 rings, we 
observed that there is a critical radius
 at which the 
trivial (or commonly used) spin representation becomes higher in 
energy than the non-trivial (twisted) spin 
representation. \ This twisted representation should correspond
to the trivial representation on a Mobius band. \ The critical 
radius is estimated to be around $10 \; nm$ for a clean 
conductor. \ The typical ring sizes today is in the 
$100-50 \; nm$ range, but it is expected that much 
smaller sizes will be available in the future \cite{saitoh}. \ 
Therefore, we claim that at these small scales the spin in the 
Mobius band and in the ring should 'behave' the same way, e.g., 
as it interacts with an external magnetic field or in 
a ferromagnetic material. \ However, before a physical analysis 
of our claim
is possible, a consistent 
 definition of spin structures 
 in non-orientable 
manifolds is needed.

\  Spin is an inherently relativistic effect of the electron and
follows from requiring Lorentz-invariance of the Schrodinger equation
\cite{dirac}. \ The relativistic treatment of the spin is not 
necessary but it considerably 
simplifies the formal discussion.

\ In this paper, we study in some detail, the different spin structures 
that
are possible in non-trivial geometries. \ For non-orientable manifolds, 
the spin group is extended to 
a  pin group where parity is violated similar to the extension of 
the special rotation group 
 $\mathbf{SO}(3)$ to the full rotation group  
$\mathbf{O}(3)$. \ In this latter case, 
the $\mathbf{Pin}(3)$ group double covers $\mathbf{O}(3)$ with 
the group $\mathbf{Spin}(3)$ being 
a connected component of $\mathbf{Pin}(3)$ which double covers 
the connected component of 
$\mathbf{O}(3)$, i.e., $\mathbf{SO}(3)$. \ So far only 
Ref. \cite{petry} 
makes use of pin structures to give a viable alternative explanation of 
a physical theory such as superconductivity. \ In this latter work, 
cooper pairs can be substituted for a more geometrical interpretation
 which is the existence of a two-inequivalent spin structures in a 
ring.  \ Hence in any non-trivial geometry, knowing the number 
of inequivalent pin structures is important to know 
before  writing a Lagrangian for the dynamics \cite{shulman2}. 

 \ For the 
sake of generality, we will allow even time non-orientable 
manifolds to be part of the discussion.  \ The work 
in the literature that covers questions
related to the existence and the counting of the different Pin
structures is mostly recent. \ A comprehensive introduction  to 
Pin group structures
can be found in Refs.~\cite{dabrowski} 
and \cite{cecile}.

\ Pin groups first appeared in Ref.\cite{atiyah} and were derived from 
Clifford algebras in relation to the K-theory 
of vector bundles. \ After that  Karoubi \cite{karoubi} studied the 
obstructions to these Pin structures within fiber bundles theory 
and hence was confined to Pin structures that are only derivable from 
a Clifford algebra.  
\  As is well known, there are
eight non-isomorphic \textbf{Z}$_{2}$ extensions of the full 
Lorentz group \cite{ebner}. \
This is a direct consequence of the disconnectedness of 
the full Lorentz group and hence there
is no unique universal two-cover as is 
the case for the special orthocronous Lorentz 
group. 

\bigskip

\ The paper is organized as follows. In section II, we review the results on
the orientable case and set the notation for the rest of  
the paper. In section III we
introduce the Pin group through the  Clifford algebra. In 
section IV we
introduce a new definition for equivalence among Pin 
structures that works in non-orientable spaces  and study its
meaning on the level of representations of the Pinor field. Section V
addresses the question of the counting of the 
inequivalent Pin structures
defined on a non-orientable manifold in 
 light of the new definition. \ In section VI, we apply 
the results to 
two cases;  
a nano-circuit that has the geometry of a torus and a 
 non-orientable de Sitter space.\ For the nanocircuit we argue 
that it is 
the nontrivial spin representation that must be used instead of the 
trivial one. \ In 
 section VII, we summarize our results.

\section{SPIN STRUCTURES}

\ In what follows, it is assumed 
 that we are dealing with manifolds with metrics that have 
the  signature $\left(+,-,-,-\right) $ and we will not address  
any questions 
that are dependent
on the metric propre. \ A good review on the mathematics involved here can 
be found in ref. \cite{eguchi}. \ In this section, we review the
spin structures on orientable manifolds and set the notation for the
rest of the paper.

For an orientable manifold $\mathcal{\mathbf{B}}$, such as
the  de Sitter space $\mathbf{R}\times \mathbf{S}%
^{3}$, a Spin structure exists whenever the second Stiefel-Whitney class
vanishes \cite{milnor}.\ This is the same as saying 
that the transition functions
of the Lorentzian frame bundle lift up to new transition functions with
values in the $\mathbf{Spin}(1,3)$ group.  The number of inequivalent 
Spin structures
is given by the number of classes in ${H}^{1}({\mathcal{\mathbf{B}}},
\mathbf{Z}_{2})$.  \ It is well known that $\mathbf{Spin}(1,3)$ is a
double cover for the Lorentz group $\mathbf{SO}_{0}(1,3)$, which is 
the connected
component of the identity of the orthogonal group $\mathbf{O}(1,3)$.  This double
covering induces some restrictions on the transition functions of a Spin
structure.

For a manifold to be orientable, the $\mathbf{O}(1,3)$ group structure of the frame
bundle should be reducible to $\mathbf{SO}(1,3)$ by choosing an orientation.  Since
the sequence

\begin{equation}
1\longrightarrow \mathbf{SO}(1,3)\longrightarrow \mathbf{O}(1,3)\longrightarrow
\mathbf{Z}_{2}\longrightarrow 1
\end{equation}

\noindent is a short exact sequence, we get the following long exact sequence \cite{hirzb},

\begin{eqnarray}
0 &\longrightarrow & H^{0}({\mathcal{\mathbf{B}}},\mathbf{SO}(1,3))\longrightarrow H^{0}({\mathcal{\mathbf{B}}}
,\mathbf{O}(1,3))\longrightarrow H^{0}({\mathcal{\mathbf{B}}},\mathbf{Z}_{2}) \nonumber \\
&\longrightarrow &H^{1}({\mathcal{\mathbf{B}}},\mathbf{SO}(1,3))\longrightarrow H^{1}({\mathcal{\mathbf{B}}}
,\mathbf{O}(1,3))\longrightarrow H^{1}({\mathcal{\mathbf{B}}},\mathbf{Z}_{2}).
\end{eqnarray}

\noindent Recalling that $H^{1}({\mathcal{\mathbf{B}}},\mathbf{SO}(1,3))$ is the set of equivalence classes of
Principal $\mathbf{SO}(1,3)$-bundles, then ${\mathcal{\mathbf{B}}}$ is orientable if and only if $\left( iff \right)$ the last map is
null.  This map is by definition the first Stiefel-Whitney class 
$\omega_{1}$. Similarly,
the short exact sequence

\begin{equation}
1\longrightarrow \mathbf{Z}_{2}\longrightarrow \mathbf{Spin}(1,3)\longrightarrow
\mathbf{SO}(1,3)\longrightarrow 1
\end{equation}
induces a long exact sequence, 
\begin{equation}
...\longrightarrow H^{1}({\mathcal{\mathbf{B}}},\mathbf{Z}_{2})\longrightarrow H^{1}({\mathcal{\mathbf{B}}
},\mathbf{Spin}(1,3))\longrightarrow H^{1}({\mathcal{\mathbf{B}}},\mathbf{SO}(1,3))\longrightarrow H^{2}({
\mathcal{\mathbf{B}}},\mathbf{Z}_{2}).
\end{equation}
Hence ${\mathcal{\mathbf{B}}}$ has a spin structure 
iff the last map is null. This
latter map is the second Stiefel-Whitney class $\omega _{2}(
{\mathcal{\mathbf{B}}})$ and hence it is the obstruction to 
a spin structure on the space $\mathbf{B}$. \ For non-orientable
spaces, $w_1 \neq 0$, we need to establish similar sequences 
to the respective groups and this will lead us naturally to the 
pin group.


\section{PINOR STRUCTURES}

\begin{figure}[ht]
  \begin{center}
 \mbox{\epsfig{file=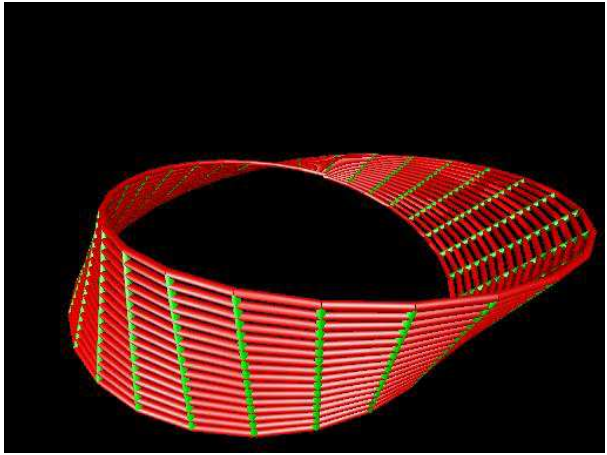,height=6 cm}} 
  \end{center}
  \caption{{{The Mobius band. }}}
\label{mobius}
\end{figure}

Pinor groups are better understood from Clifford algebras \cite{atiyah}. Given a
vector space $\mathcal{V}$ of dimension $n=4$ that is tangent to a point $p\in 
{\mathcal{\mathbf{B}}}$ and $g(x,y)$  a bilinear non-degenerate quadratic form on $\mathcal{V}$
associated with the metric $(+,-,-,-)$. Let $T(\mathcal{V})=\sum_{i=0}^{\infty
}T^{i}(\mathcal{V})$, where 
\begin{equation}
T^{i}(\mathcal{V})=\underset{i-times}{\underbrace{\mathcal{V}\times ...\times
\mathcal{V}}}, 
\end{equation}
is the tensor algebra of $\mathcal{V}$.
The set $\mathcal{I}$ generated by the set$\{x\otimes x-g(x,x),x\in \mathcal{V}\}$ is an ideal
of $T(\mathcal{V}).$ \ The quotient space 
\begin{equation}
Cl(\mathcal{V},Q)=\frac{T(\mathcal{V})}{\mathcal{I}}
\end{equation}
is the Clifford algebra of the vector space $\mathcal{V}$ equipped with the quadratic
form $Q(x)=g(x,x).$ \ Clearly, if $Q(x)=0, Cl(\mathcal{V})$ is simply the Grassmann
algebra of forms on $\mathcal{V}$. The dimension of $Cl(\mathcal{V},Q)$ is 2$^{4}.$ \ The
multiplication in this algebra is induced by the tensor product in $T(\mathcal{V}).$ \
Let $\left\{ e_{0},e_{1},e_{2},e_{3}\right\} $ be an orthonormal basis for $%
\mathcal{V} $. \ Then, the following vectors of $Cl(\mathcal{V},Q)$, 
\begin{eqnarray*}
&&e_{0},e_{1},e_{2},e_{3} \nonumber \\
&&e_{0}e_{1},e_{0}e_{2},e_{0}e_{3},e_{1}e_{2},e_{1}e_{3},e_{2}e_{3} \nonumber \\
&&e_{0}e_{1}e_{2},e_{0}e_{1}e_{3},e_{0}e_{2}e_{3},e_{1}e_{2}e_{3} \nonumber \\
&&e_{0}e_{1}e_{3},1
\end{eqnarray*}
form a basis for the algebra. \ Moreover, since 
\begin{eqnarray}
Q(e_{\mu }+e_{\nu }) &=&(e_{\mu }+e_{\nu })\otimes (e_{\mu }+e_{\nu }) \nonumber \\
&=&e_{\mu }\otimes e_{\mu }+e_{\mu }\otimes e_{\nu }+e_{\nu }\otimes e_{\mu
}+e_{\nu }\otimes e_{\nu },
\end{eqnarray}
we have, on the level of algebra,
\begin{eqnarray}
e_{\mu }e_{\nu }+e_{\nu }e_{\mu } &=&0\qquad \text{\ \ \ \ \ \ \ if }\mu
\neq \nu \nonumber \\
\ e_{\mu }^{2} &=&2e^{\mu }e_{\mu }\qquad \mu =0,1,2,3 .
\end{eqnarray}

\noindent Clearly Dirac $\gamma _{\mu }$ matrices form a representation of this
algebra ( Majorana representation). \ The involution $x\rightarrow -x$ in $\mathcal{V}$
extends naturally to an involution $\alpha $ of the algebra which in turn
induces a $\mathbf{Z}_{2}$ grading of $Cl(\mathcal{V},Q)$, i.e.,

\begin{equation}
Cl(\mathcal{V},Q)=Cl^{+}(\mathcal{V},Q)+Cl^{-}(\mathcal{V},Q),
\end{equation}
where $Cl^{+}(\mathcal{V},Q) \left( Cl^{-}(\mathcal{V},Q)\right)$ contains the even 
(odd) elements of the
algebra.

A norm function can be defined on $Cl(\mathcal{V},Q)$ by first defining conjugation on
the generators:

\begin{equation}
(e_{i_{1}}....e_{i_{p}})^{\ast }=e_{i_{p}}....e_{i_{1}} \: .
\end{equation}
Then the norm of $x$ is defined by

\begin{equation}
\left\| x\right\| ^{2}=x\alpha (x^{\ast }) .
\end{equation}

Let $Cl^{\ast }(\mathcal{V},Q)$ be the subset of all invertible elements of $Cl(\mathcal{V},Q).$
\ Actually in our case where the dimension is even we might as well define
the norm without the inversion $\alpha .$ \ The Clifford group $\Gamma (\mathcal{V},Q)$
is the subgroup of $Cl^{\ast }(\mathcal{V},Q)$ \ defined by

\begin{equation}
\Gamma (\mathcal{V},Q)=\left\{ x\in Cl^{\ast }(\mathcal{V},Q): v \in \mathcal{V},\alpha (x)vx^{-1}\in
\mathcal{V}\right\}.
\end{equation}

\bigskip Given $x\in \Gamma (\mathcal{V},Q)$, then it can be represented by an
orthogonal transformation

\begin{eqnarray}
\rho (x) &:&\mathcal{V}\rightarrow \mathcal{V} \nonumber \\
e &\rightarrow &\alpha (x)ex^{-1}
\end{eqnarray}

\noindent If we were to drop $\alpha $ from the 
definition, the map $\rho $ fails to
be a representation in the odd dimensional case. \ Finally, 
the {\it Pin group} is
the subgroup of the Clifford group with elements of modulus one,

\begin{equation}
\mathbf{Pin}(\mathcal{V},Q)=\left\{ x\in \Gamma (\mathcal{V},Q):\qquad ||x||=1\right\} .
\end{equation}
This group doubly covers the orthogonal group $\mathbf{O}(1,3)$. The sequence 
\begin{equation}
1\longrightarrow \mathbf{Z}_{2}\longrightarrow \mathbf{Pin}(1,3)\longrightarrow
\mathbf{O}(1,3)\longrightarrow 1
\end{equation}
is then a short exact sequence. We say that $\mathbf{Pin}(1,3)$ is a $\mathbf{Z}_{2}-$%
extension of $\mathbf{O}(1,3)$.

\ Finally, it is important to realize that not all pin groups
are derived from a Clifford algebra.\  Chamblin \cite{chamblin} 
studied  
the obstructions to
non-Cliffordian Pin structures. \ Starting from Dabrowski's 
semidirect product formula for the Pin group ~\cite
{dabrowski}, \ Chamblin found an obstruction to Pin 
structures through the use of Sheaf
cohomology \cite{hirzb}. The starting short exact sequence 
that was fundamental
to his construction  is however not correct. \ This is easily
seen by applying the second homomorphism theorem. Dabrowski's formula is 
\begin{equation}
\mathbf{Pin}^{a,b,c}(p,q)=\frac{\mathbf{Spin}_{0}(p,q)\odot \mathbf{C}^{a,b,c}}{\mathbf{Z}_{2}},
\end{equation}
where $\mathbf{C}^{a,b,c}$ stand for the discrete group of order 8 that is a double
covering of the group $\mathbf{G}=\left\{ 1,T,P,T P\right\} $.  The rotation group $
\mathbf{O}(p,q)=\mathbf{SO}(p,q)\odot G$ is  double covered by the Pin group and there are
eight non-isomorphic such groups. \ This latter eight is unrelated to the
 order of the group  $\mathbf{C}^{a,b,c}$ .  The group 
$\mathbf{G}$ is isomorphic to $\mathbf{Z}_{2}\otimes \mathbf{Z}_{2}.$
The group $\mathbf{C}^{a,b,c}$ is isomorphic to either of the following groups.  The
quaternion group $\mathbf{Q}$, the dihedral group $\mathbf{D}_{4}$, the group $\mathbf{Z}_{2}\times
\mathbf{Z}_{2}\times \mathbf{Z}_{2}$ and the group $\mathbf{Z}_{2}\times \mathbf{Z}_{4}$.  The a, b 
and c indexes stand for the
signs of the squares of the elements of the cover $T, P$ and $P T.$  For
example in the quaternion case we 
can map $i$ to $T$, $j$ to $P$ and $k$ to $
T P$.  In this case $a=b=c=-$, and so on. \ In the rest of the paper, we 
will be only interested in determining the degrees of freedom on 
the spin structures for a given topology. We hope to address the 
question of obstructions in the future.

\bigskip

\section{WEAKLY-EQUIVALENT PIN STRUCTURES}

In this section, we introduce a new definition for equivalence 
among (s)pin structures that takes into account the
possible non-orientability 
of the physical space of the system. \ The definition is suggested such
that the non-orientability is linked to the orientable cover of the
space. \ This is in analogy with relating non-simply connected spaces
to their simply connected covers.

\subsection{Pinor-Frames}

Let $\mathcal{\mathbf{B}}$ a manifold with a metric of 
signature $(+,-,-,-)$ and covered by a
simple cover $\left\{ U_{i}\right\} ,\Gamma $ is the $\mathbf{Pin}(1,3)$ 
group and $\mathbf{G}$
is the orthogonal group $\mathbf{O}(1,3)$. $(\mathcal{\mathcal{\mathbf{P}}},\pi ,\mathcal{\mathbf{B}},\mathbf{G})$ is a principal bundle. $(
\widetilde{\mathcal{\mathbf{P}}},\widetilde{\pi },\mathcal{\mathbf{B}},\Gamma )$ is the principal bundle induced
by the double covering 
\begin{equation}
\rho :\Gamma \rightarrow \mathbf{G} \: .
\end{equation}
There is a bundle map $\Phi $ between\ $\widetilde{\mathcal{\mathbf{P}}}$\ and $\mathcal{\mathbf{P}}$\ such that 
\begin{eqnarray}
\Phi \pi &=&\widetilde{\pi } \nonumber \\
\Phi (\upsilon \cdot \gamma ) &=&\Phi (\upsilon )\Phi (\gamma )
\end{eqnarray}
for $\upsilon \in $\ $\widetilde{\mathcal{\mathbf{P}}}$ and $\gamma \in \Gamma $ .

Consider now two $\Gamma -$structures  $\widetilde{\mathcal{\mathbf{P}}}$ and  $\widetilde{\mathcal{\mathbf{P}}^{\prime}}$ over $\mathcal{\mathbf{P}}$
that differ only through an automorphism $\Psi $\ of $\Gamma $, in other
words, we have a bundle isomorphism $\Theta $\ such that

\begin{equation}
\begin{array}{llll}
\Theta : & \ \widetilde{\mathcal{\mathbf{P}}} & \rightarrow & \ \widetilde{\mathcal{\mathbf{P}}}^{\prime } \\ 
& \Phi \searrow &  & \swarrow \Phi ^{\prime } \\ 
&  & \mathcal{\mathbf{P}} & 
\end{array}
\end{equation}
commutes and 
\begin{equation}
\Theta (\upsilon \cdot \gamma )=\Theta (\upsilon )\cdot \Psi (\gamma )\text{
\ \ \ \ \ \ }\upsilon \in \widetilde{\mathcal{\mathbf{P}}},\gamma \in \Gamma \: .
\end{equation}
$\widetilde{\mathcal{\mathbf{P}}}$ and $\mathcal{\mathbf{P}}$\ are said to be \textit{weakly-equivalent}. 
Because of the double covering, the map $\Psi $\ is involutive.

Next, we state the following definition of 
a pinor field which is a 
generalization of the usual spinor field.
{\it{A Pinor field  $\psi $ of type ($\xi ,\mathbf{Y})$ on $\mathcal{\mathbf{B}}$
 is a section of $
\widetilde{\mathcal{\mathbf{P}}}\times _{\Gamma } \mathbf{Y}$ \ with 
\begin{equation}
\xi :\Gamma \rightarrow Hom(\mathbf{Y},\mathbf{Y})
\end{equation}
and is a representation map of $\Gamma $ .}}

\subsubsection*{\textbf{Theorem I}}
{\it{The sets of Pinor fields representations in $\widetilde{\mathcal{\mathbf{P}}}$ and $\widetilde{\mathcal{\mathbf{P}}
}^{\prime }$\ are $\Psi -$\ related.}}

\bigskip
To prove this we first
 take note of the fact that a section $\sigma $ of $
\widetilde{\mathcal{\mathbf{P}}}\times _{\Gamma }Y$\ \ can be represented as a map 
\begin{eqnarray}
S &:&\widetilde{\mathcal{\mathbf{P}}}\rightarrow \mathbf{Y} \nonumber \\
S(\widetilde{u}) &=&\widetilde{u}^{-1}(\sigma (x))
\end{eqnarray}
such that 
\begin{equation}
S(\widetilde{u}\cdot \gamma )=\gamma ^{-1}\cdot S(\widetilde{u})
\end{equation}
Here we have used the fact that an element $\widetilde{u}$ of $\widetilde{\mathcal{\mathbf{P}}}$
 represents a map from $\mathbf{Y}$ to $Y_{x}$ .  Moreover, the bundle map $\Theta $
 can be taken to be the identity map on fibers, i.e., 
\begin{equation}
\Theta \equiv 1:Y_{x}\rightarrow Y_{x} 
\end{equation}

\bigskip

Let $\left\{ \gamma _{ij}\right\} $ and \ $\left\{ \gamma _{ij}^{\prime
}\right\} $\ be the transition functions of $\widetilde{\mathcal{\mathbf{P}}}$\ and $\widetilde{
\mathcal{\mathbf{P}}}^{\prime }$\ , respectively.  Then it is obvious that the following
diagram commutes: 
\begin{equation}
\begin{tabular}{llllll}
$\gamma _{ji}:$ & $U_{i}\smallfrown U_{j}$ &  & $\rightarrow $ &  & $\Gamma $
\\ 
&  &  &  & $\Psi $ & $\downarrow $ \\ 
&  & $\gamma _{ji}^{\prime }$ & $\searrow $ &  &  \\ 
&  &  &  &  & $\Gamma $%
\end{tabular}
\end{equation}
i.e., 
\begin{equation}
\Psi (\gamma _{ji})=\gamma _{ji}^{\prime } \: .
\end{equation}
Therefore given a section $\sigma _{i} : U_{i}\rightarrow \widetilde{\mathcal{\mathbf{P}}}\times
_{\Gamma }\mathbf{Y}$, there is a corresponding one $\sigma _{i}^{\prime
}~: U_{i}\rightarrow \widetilde{\mathcal{\mathbf{P}}}^{\prime }\times _{\Gamma }\mathbf{Y}$ such that 
\begin{eqnarray}
\sigma _{i}(x) &=&[x,y]=[x\cdot \gamma ,\gamma ^{-1}\cdot y] \nonumber \\
\sigma _{i}^{\prime }(x) &=&[x,y]=[x\cdot \Psi (\gamma ),\Psi (\gamma
^{-1})\cdot y]
\end{eqnarray}
An element $u$\ of $\widetilde{\mathcal{\mathbf{P}}}$ can be represented by an equivalence
class $[i,x,\gamma ] = [j,x,\gamma _{ji}\gamma ]$  with a similar expression
for $u^{\prime }$ of $\widetilde{\mathcal{\mathbf{P}}}^{\prime }$\ with $\gamma _{ji}$ 
replaced by $\Psi (\gamma _{ji})$ .  This is equivalent to saying that $u$
 is a map 
\begin{equation}
\mathbf{Y}\rightarrow Y_{x}
\end{equation}
with 
\begin{equation}
u(y)=\varphi _{i}(x,\gamma \cdot y) \: .
\end{equation}
The $\varphi _{i}$'s are the local trivializations of $\widetilde{\mathcal{\mathbf{P}}}$. 
For the cross sections we have the following diagram 
\begin{equation}
\begin{array}{ccccc}
S: & \widetilde{\mathcal{\mathbf{P}}} & \rightarrow &  & \mathbf{Y} \\ 
\Theta & \downarrow &  & \nearrow & S^{\prime } \\ 
& \widetilde{\mathcal{\mathbf{P}}}^{\prime } &  &  & 
\end{array}
\end{equation}
so $S(u)=S^{\prime }(\Theta (u))=S^{\prime }(u^{\prime })$.  This in turn
implies that

\begin{eqnarray}
S^{\prime }(u^{\prime }\cdot \gamma ^{\prime }) &=&\gamma ^{^{\prime
}-1}\cdot S^{\prime }(u^{\prime }) \nonumber \\
&=&\Psi (\gamma ^{-1})\cdot S(u) \: .
\end{eqnarray}
If $\xi $ is a representation of $\Gamma $\ , the above relation trivially
extends to 
\begin{equation}
S^{\prime }(u^{\prime }\cdot \xi (\gamma ^{\prime }))=\Psi (\xi (\gamma
^{-1}))\cdot S(u) \: .
\end{equation}
We conclude that a pinor field defined on $\mathcal{\mathbf{B}}$ is a quantity independent of
the action group within an isomorphism.
\bigskip

\subsection{On frame-Bundles}

Let as above $\mathcal{\mathbf{B}}$ be a 
non-orientable manifold and $\mathcal{\mathbf{B}}^c$\ its
 {\it orientable double cover}.  Here, we propose to treat the question of what
happens if the frame bundle $\overline{\mathcal{\mathbf{P}}}$, with 
group of action taken to be 
$\mathbf{O}_{+}^{\uparrow }(1,3)$\ , is pushed forward with the 
covering map $p : \mathcal{\mathbf
{B}}^c\rightarrow \mathcal{\mathbf{B}}$.  We 
take  $\{U_{i}\}$\ as the cover of $\mathcal{\mathbf{B}}$.

First it should be realized that the transition functions $\overline{g_{ji}}
(x^{\prime })$\ of $\overline{\mathcal{\mathbf{P}}}$ \ are the same as those of the tangent
bundle $T\mathcal{\mathbf{B}}^c$ .  By definition, these transition functions are
the Jacobian of the transition functions of the coordinate functions: 
\begin{equation}
\varphi _{i}^{\prime }:E_{i}\rightarrow V_{i}=p^{-1}(U_{i}) \: ,
\end{equation}
where $E_{i} \subset \mathbf{R^{4}}$ .  The transition functions 
$t_{ji}(x^{\prime })$ of  the chart 
$(V_{i},\varphi _{i})$ are given by 
\begin{equation}
t_{ji}^{\prime } = \varphi _{i}^{\prime -1}\varphi _{i}^{\prime}:
  E_{i}\smallfrown E_{j}\rightarrow E_{i}\smallfrown E_{j} \: .
\end{equation}
Therefore the transition functions $a_{ji}(x^{\prime })$ of $T\mathcal{\mathbf{B}}^c$
 are given by : 
\begin{eqnarray}
a_{ji}^{\prime } &:&V_{i}\smallfrown V_{j}\rightarrow \mathbf{O}_{+}^{\uparrow }(1,3)
 \nonumber \\
x^{\prime } &\longmapsto &a_{ji}^{\prime }(x^{\prime })=J(t_{ji}^{\prime
})|_{\varphi _{i}^{-1}(x^{\prime })} \: .
\end{eqnarray}
Hence we have 
\begin{equation}
\overline{g_{ji}}(x^{\prime })\ =a_{ji}^{\prime }(x^{\prime }) \: .
\end{equation}
This can be easily shown through the coordinate functions.

Now, we describe in more detail the map $p:\mathcal{\mathbf{B}}^c$\ $\rightarrow \mathcal{\mathbf{B}}$\ .
The orientable double cover is defined as follows.  The Jacobian of the
transition functions of $\mathcal{\mathbf{B}}$, $a_{ji}(x)$, that corresponds to $
a_{ji}^{\prime }$ in $\mathcal{\mathbf{B}}^c$ are defined similarly.  From these
transition functions, we can form 1-cochains $\theta $  
\begin{eqnarray}
\theta _{ij} &:&U_{i}\smallfrown U_{j}\rightarrow \mathbf{Z}_{2} \nonumber \\
x &\longmapsto &\theta _{ji}(x)=\det [J(a_{ji})|_{\varphi _{i}^{-1}(x)}].
\end{eqnarray}
Therefore to each point $x\in \mathcal{\mathbf{B}}$\ , we can associate to it two points $(x,1)$
and $(x,-1)$ where $\pm 1$\ is the values of $\theta _{ij}(x)$ .  The
manifold $\mathcal{\mathbf{B}}^c$\ is the set of all these points.  First we note
that $\mathcal{\mathbf{B}}^c$ is connected.  A curve in $\mathcal{\mathbf{B}}^c$ that
connects  $(x,1)$ to $(x,-1)$ can be given through the unique 
lifting of a non-orientable
closed loop at $x$\ \cite{spanier}. $\mathcal{\mathbf{B}}^c$\ is 
also orientable since
lifting the $\theta _{ij}$'s will give the determinant of the Jacobian of
the $a_{ij}$'s.  The manifold $\mathcal{\mathbf{B}}^c$\ can in fact be interpreted as
a bundle structure over $\mathcal{\mathbf{B}}$ with fiber and group $\mathbf{Z}_{2}$ . 
Diagrammatically we have 
\begin{equation}
\begin{tabular}{lll}
$\overline{\mathcal{\mathbf{P}}}$ &  &  \\ 
$\downarrow \overline{\pi }$ &  &  \\ 
$\mathcal{\mathbf{B}}^c$ & $\rightarrow $ & $\mathcal{\mathbf{B}}$ \\ 
& $p$ & 
\end{tabular}
\end{equation}
Therefore for the principal bundle $(\mathcal{\mathbf{B}}^c,p,\mathcal{\mathbf{B}},\mathbf{Z}_{2})$ the local
trivializations are given by 
\begin{equation}
\phi _{i}:U_{i}\times \mathbf{Z}_{2}\rightarrow p^{-1}(U_{i})=V_{i}
\end{equation}
and the transition functions $g_{ji}(x):\mathbf{Z}_{2}\rightarrow \mathbf{Z}_{2}$\ are
elements of $\mathbf{Z}_{2}$\ , they act as a permutation group of the points that
cover $x$ .  

Now define a new set $P=\smallsmile _{x\in \mathcal{\mathbf{B}}}P_{x}$\ where $P_{x}=\overline{\mathcal{\mathbf{P}}
}_{x_{1}^{\prime }}\smallsmile $\ $\overline{\mathcal{\mathbf{P}}}_{x_{2}^{\prime }}$\ if $
p(x_{1}^{\prime })=$\ $p(x_{2}^{\prime })=x$ and distinct.  We claim that
the map 
\begin{eqnarray}
p_{\ast } &:&\mathcal{\mathbf{P}}\rightarrow \mathcal{\mathbf{B}} \nonumber \\
p_{\ast }(u) &=&x\text{ where }u\in \overline{\mathcal{\mathbf{P}}}\text{ and }p(\overline{\pi }
(u))=x
\end{eqnarray}
induces a bundle structure on $\mathcal{\mathbf{B}}$.  Therefore we expect the following
diagram to commute 
\begin{equation}
\begin{tabular}{llll}
$\Lambda :$ & $\overline{\mathcal{\mathbf{P}}}$ & $\rightarrow $ & $\mathcal{\mathbf{P}}$ \\ 
& $\overline{\pi }\downarrow $ &  & $\downarrow p_{\ast }$ \\ 
& $\mathcal{\mathbf{B}}^c$ & $\rightarrow$ & $\mathcal{\mathbf{B}}$ \\ 
&  & $p$ & 
\end{tabular}
\end{equation}
besides taking fiber to fiber, the map $\Lambda $ should respect the action
of the respective groups in both manifolds: 
\begin{equation}
\begin{tabular}{lll}
$\overline{\mathcal{\mathbf{P}}}\times \mathbf{O}_{+}^{\uparrow }(1,3)$ & $\rightarrow $ & $\overline{\mathcal{\mathbf{P}}}$
\\ 
$\Lambda \downarrow \tau $ &  & $\downarrow $ \\ 
$\mathcal{\mathbf{P}}\times \mathbf{G}$ & $\rightarrow $ & $\mathcal{\mathbf{P}}$%
\end{tabular}
\end{equation}
\begin{equation}
\Lambda (u\cdot \lambda )=\Lambda (u)\cdot \tau (\lambda )
\end{equation}
We would like to find out the group structure $\mathbf{G}$ of this bundle.  Assuming
that $\mathcal{\mathbf{B}}$ has a metric with signature $(+,-,-,-)$ then the functions $
a_{ji}(x)$ defined above are in $\mathbf{O}(1,3)$ .  Since $\mathcal{\mathbf{B}}$ is not orientable
then $\det a_{ji}(x)=\pm 1$.  The elements  that cover  $x$ differ by
an element in $\mathbf{Z_{2}}$ which can be represented by $T$ or $P$  
or any other element with determinant -1 and involutive.  The first matrix
is related to time non-orientability, the second to space non-orientability.
 These elements describe global actions on the manifold.
By construction, the fiber $Y_{x}\simeq \mathbf{O}_{+}^{\uparrow }\smallsmile
\mathbf{O}_{+}^{\uparrow }$. So from the above discussion we should expect that the
two copies differ by an element of determinant -1.  In fact let $a$ be such an
element with $a^{2}=1$. Using the map $\mathbf{Z}_{2}\times $ $\mathcal{\mathbf{B}}^c\rightarrow \mathcal{\mathbf{B}}$ , a
well defined multiplication,  then we can write 
\begin{equation}
x_{2}^{\prime }=a\cdot x_{1}^{\prime } \: ,
\end{equation}
where the local coordinates are used , that is 
\begin{eqnarray}
x_{1}^{\prime } &:&V_{i}\rightarrow E_{i} \nonumber \\
x_{2}^{\prime } &:&V_{j}\rightarrow E_{j} \: .
\end{eqnarray}
On the manifold $\mathcal{\mathbf{B}}$ these charts get projected to a single chart around $x$: 
\begin{eqnarray}
x_{1}^{\prime } &:&\text{ \ \ \ }y_{1}:U_{i}\rightarrow E_{i} \nonumber \\
x_{2}^{\prime } &:&\text{ \ \ \ }y_{2}:U_{j}\rightarrow E_{j} \: .
\end{eqnarray}

\noindent Therefore, the 
frames \ $u_{1}(x_{1}^{\prime })$ and $u_{2}(x_{2}^{\prime })$
 become two frames at the same point $x$.  They are related through the
transformation $\frac{\partial y_{1}^{\mu }}{\partial y_{1}^{\upsilon }}
|_{x} $.  But $y_{1}^{\mu }(x)=x_{1}^{\prime \mu }(x)=a_{\epsilon
}^{-1}z_{1}^{^{\prime }\varepsilon }(x_{2}^{\prime })$ and $y_{2}^{\upsilon
}(x)=y_{2}^{^{\prime }\upsilon }(x_{2}^{\prime })$, this implies that 
\begin{equation}
\frac{\partial y_{1}^{\mu}}{\partial y_{2}^{\nu}} \; = 
\; \left ( a^{-1} \right )_{\epsilon}^{\mu} \;  
\frac{\partial {z_{2}^{\prime }}^{\epsilon}}{\partial {y_{2}^{\prime }}^{\nu} 
\; \mid _{x_{2}^\prime} }\: .
\end{equation}
Since $\frac{\partial z_{2}^{\prime }}{\partial y_{2}^{\prime }}\in
\mathbf{O}_{-}^{\uparrow }(1,3)$, we conclude that the transition functions of $\mathcal{\mathbf{P}}$
are in $\mathbf{O}_{+}^{\uparrow }(1,3)\smallsmile a\cdot \mathbf{O}_{+}^{\uparrow }(1,3)$ .
 There are two possible choices either $a\in \mathbf{O}_{-}^{\uparrow }(1,3)$  or  
$a\in \mathbf{O}_{-}^{\downarrow }(1,3)$ . Depending 
on which element $a$ we choose, we end up
with different actions on $\mathcal{\mathbf{P}}$ which are equivalent.  So the group $\mathbf{G}$ can be
either $\mathbf{O}_{+}^{\uparrow }(1,3)\smallsmile $ $\mathbf{O}_{-}^{\uparrow }(1,3)$  or $
\mathbf{O}_{+}^{\uparrow }(1,3)\smallsmile $ $\mathbf{O}_{-}^{\downarrow }(1,3)$.  Hence, if
we have started with $\mathbf{O}_{+}^{\uparrow }(1,3)\smallsmile $ 
$\mathbf{O}_{+}^{\downarrow }(1,3)$ as the group of symmetry of $\overline{\mathcal{\mathbf{P}}}$ , we
would have obtained $\mathbf{O}(1,3)$ as the group of action of $\mathcal{\mathbf{P}}$. On the level
of fibers, the map $\Lambda$
 is easily seen to be a 2-1 map similar to $p$
by construction.

\section{WEAKLY-INEQUIVALENT PIN STRUCTURES }

In the following , the topological group $\Gamma $\ can be taken to be $
\mathbf{Pin}(p,q)$ and the group $\mathbf{G}$
 the orthogonal group $\mathbf{O}(p,q)$.

Let $(\mathcal{\mathbf{P}}$,$\pi ,\mathcal{\mathbf{B}},\mathbf{G})$\ be a Principal bundle.  Let $\Gamma $ be a double
covering for $\mathbf{G}$.  Then we have the following exact sequence: 
\begin{equation}
1\rightarrow K\rightarrow \Gamma \rightarrow \mathbf{G}\rightarrow 1 \: ,
\end{equation}
where $K=\mathbf{Z}_{2}$.  And let 
\begin{equation}
\rho :\Gamma \rightarrow \mathbf{G}
\end{equation}
be the 2-1 map.  We denote by $g_{ji} : U_{i}\cap U_{j} \rightarrow \mathbf{G}$
 the transition functions of the bundle $\mathcal{\mathbf{P}}$.\ 
 The principal bundle $(\widetilde{\mathcal{\mathbf{P}}}$,$\widetilde{\pi },\mathcal{\mathbf{B}},\mathbf{G})$ is called a 
{\it $\Gamma$
-structure on $\mathcal{\mathbf{P}}$} iff there is a bundle map 
\begin{eqnarray}
\Phi &:&\widetilde{\mathcal{\mathbf{P}}}\rightarrow \mathcal{\mathbf{P}}  \nonumber \\
\Phi (\widetilde{u}\cdot \gamma ) &=&\Phi (\widetilde{u})\cdot \rho (\gamma )
\: \: \: \: \text{ for }\widetilde{u}\in \widetilde{\mathcal{\mathbf{P}}}\text{\ and }\gamma \in \Gamma
\end{eqnarray}
and the functions $g_{ji}$ can be lifted to
 the transition functions of $
\widetilde{\mathcal{\mathbf{P}}}$\ , i.e., we have 
\begin{equation}
\begin{tabular}{lll}
$\gamma _{ji}$ &  & $\Gamma $ \\ 
& $\nearrow $ & $\downarrow $ \\ 
$U_{ji}$ & $\overset{g_{ji}}{\rightarrow }$ & $\mathbf{G}$
\end{tabular}
\end{equation}

\begin{eqnarray}
\rho (\gamma _{ji}) &=& g_{ji} \nonumber  \\
\gamma _{ij}\gamma _{kj}^{-1}\gamma _{ki} &=&1 \: .
\end{eqnarray}

The last relation is the cocycle condition.  This enables us to define a
\^{C}ech-Cohomology on $\mathcal{\mathbf{B}}$ with coefficients in  $\Gamma $: 
\begin{eqnarray}
d &:&C^{1}\longrightarrow C^{2} \nonumber \\
(d\gamma )(ijk) &=&\gamma _{jk}\gamma _{ik}^{-1}\gamma _{ij} \: .
\end{eqnarray}

Hence $\left[ \gamma _{ij}\right] \in H^{1}(\mathcal{\mathbf{B}},\Gamma )$ and $\left[ g_{ij}
\right] \in H^{1}(\mathcal{\mathbf{B}},\mathbf{G}).$  The equivalence class $\left[ \gamma _{ij}\right] 
$ is defined by 
\begin{equation}
\left[ \gamma _{ij}\right] =\{\gamma _{jk}^{\prime }:\lambda _{i}\gamma
_{ij}\lambda _{j}^{-1}\}
\end{equation}
with 
\begin{equation}
\lambda _{i}:U_{i}\rightarrow \Gamma \: .
\end{equation}
It is immediately clear from the above, that $\rho $\ induces a map $\rho
^{\ast }$ between cohomologies: 
\begin{eqnarray}
\rho ^{\ast } &:&H^{1}(\mathcal{\mathbf{B}},\Gamma )\rightarrow H^{1}(\mathcal{\mathbf{B}},\Gamma ) \nonumber \\
\rho ^{\ast }\left[ \gamma _{ij}\right] &=&[\rho \left[ \gamma _{ij}\right]
]=\left[ g_{ij}\right] \: .
\end{eqnarray}
This map is 1-1 and onto if a $\Gamma -$\ structure exists.

Now if
 we assume that the manifold $\mathcal{\mathbf{B}}$ is non-orientable.  Then $\mathcal{\mathbf{B}}$ has an
orientable double cover $\mathcal{\mathbf{B}}^c$.  Let \{$U_{1}\}$ be a cover for $\mathcal{\mathbf{B}}$
, and 
\begin{equation}
\varphi _{i}:E_{i}\rightarrow U_{i}
\end{equation}
be the coordinate maps of $\mathcal{\mathbf{B}}$ where $E_{i}\subset R^{n}$.  The transition
maps $a_{ij}$ are given by 
\begin{equation}
a_{ij}:\varphi _{j}\varphi _{i}^{-1}:E_{i}\rightarrow E_{j}\rightarrow
E_{i}\rightarrow E_{j} \: .
\end{equation}
Now let $x\in U_{i}\smallfrown U_{j}$\ and define $\theta _{ij}$\ to be the
normalized determinant of the Jacobian of the transition functions $a_{ij}:$
\begin{eqnarray}
\theta _{ij} &:&U_{i}\smallfrown U_{j}\rightarrow \mathbf{Z}_{2} \nonumber \\
\theta _{ij}(x) &=&\det J[a_{ij}(x)] \: .
\end{eqnarray}
Using the properties of the determinant, we can see that $\theta _{ij}$ is a
representative of an element of $H^{1}(\mathcal{\mathbf{B}},\mathbf{Z}_{2})$ .  We construct $
\mathcal{\mathbf{B}}^c$ through these cocycles: 
\begin{equation}
p:\mathcal{\mathbf{B}}^c\rightarrow \mathcal{\mathbf{B}} \: ,
\end{equation}
where
\begin{equation}
\mathcal{\mathbf{B}}^c=\left\{ \left( x,\theta _{ij}(x)\right) ,x\in U_{ij}\right\} \: .
\end{equation}
Therefore, $p$ induces a bundle structure.  In fact $(\mathcal{\mathbf{B}}^c
,p,\mathcal{\mathbf{B}},\mathbf{Z}_{2})$ is a principal bundle.  Note that this fiber bundle is
connected if the base space $\mathcal{\mathbf{B}}$ is connected.  If $\mathcal{\mathbf{B}}$ were orientable then $
\mathcal{\mathbf{B}}^c$ would simply be a trivial double covering.

 \ Next, we let
 $\widetilde{\mathcal{\mathbf{P}}}$\ and 
$\widetilde{\mathcal{\mathbf{P}}}^{\prime }$\ be two 
$\Gamma $%
-structures on $\mathcal{\mathbf{P}}$ and 
$\Psi $ is a group isomorphism of $\Gamma $ such that 
\begin{equation}
\begin{tabular}{lllll}
$\Psi :$ & $\Gamma $ & $\rightarrow $ &  & $\Gamma $ \\ 
& $\rho \searrow $ &  & $\swarrow \rho $ &  \\ 
&  & $\mathbf{G}$ &  & 
\end{tabular}
\end{equation}
commutes.  $\widetilde{\mathcal{\mathbf{P}}}$\ 
and $\widetilde{\mathcal{\mathbf{P}}}^{\prime }$\ 
are said to
be {\it weakly equivalent}, 
$\widetilde{\mathcal{\mathbf{P}}}$ $\simeq _{W}\widetilde{\mathcal{\mathbf{P}}}^{\prime
}$\ , iff we have the following commutative diagram 
\begin{equation}
\begin{tabular}{lllll}
$\widetilde{\mathcal{\mathbf{P}}}$ &  & $\overset{\Theta \simeq }{\rightarrow }$ &  & $
\widetilde{\mathcal{\mathbf{P}}}^{\prime }$ \\ 
& $\Phi \searrow $ &  & $\swarrow \Phi ^{\prime }$ &  \\ 
&  & $\mathcal{\mathbf{P}}$ &  & 
\end{tabular}
\end{equation}
and 
\begin{equation}
\Theta (\widetilde{u}\cdot \gamma )=\Theta (\widetilde{u})\cdot \Psi (\gamma
)\text{ \ \ \ \ \ \ }\widetilde{u}\in \widetilde{\mathcal{\mathbf{P}}},\gamma \in \Gamma \: .
\end{equation}

\noindent If $\Psi =1,$ we say the {\it equivalence is strong} 
\cite{greub}. \ We believe that weak 
equivalence
is the concept most appropriate from a physical point of view.  We 
have seen
above that such fields are $\Psi $-related, moreover this will enable us to
concern ourselves only with fields defined on orientable 
manifolds. \ Therefore 
if we are 
interested in knowing how many possible
physical fields we can have on a given manifold, it is irrelevant how we
represent the group of actions on the spinor field as long as they differ by
an isomorphism. \ We show next that
the map $\Psi $ is an involution. \ To show this, 
we observe that
the map $\rho $ is a 2-1 map.  Let $\gamma ^{1}$\ and $\gamma ^{2}$ the
two elements that cover $g$.  Since $\rho \Psi (\gamma ^{1})=g$  and  $
\rho (\gamma ^{2})=g$, we must either have $\Psi = 1$ or 
\begin{eqnarray}
\Psi (\gamma ^{1}) &=&\gamma ^{2} \nonumber \\
\text{and }\Psi (\gamma ^{2}) &=&\gamma ^{1} \: .
\end{eqnarray}
The map $\Psi $ acts as a permutation among the couple that cover a given
element in $\mathbf{G}$. \  Hence the involution follows immediately.
A result that follows from the previous statement is that
$\Psi |_{K}=1$ \: . \ This is immediate 
from the fact that $K=\mathbf{Z}_{2}$.

\subsubsection*{\textbf{Theorem II }}
{\it{Let $\widetilde{\mathcal{\mathbf{P}}},\widetilde{\mathcal{\mathbf{P}}}^{\prime }$\ be $\Gamma$ -structures on $\mathcal{\mathbf{P}}$,

\begin{enumerate}
\item  Let $\widetilde{\mathcal{\mathbf{P}}}\simeq _{W}\widetilde{\mathcal{\mathbf{P}}}^{\prime }$\ . If $
\left\{ \gamma _{ij}\right\} $ are transition functions for $\widetilde{\mathcal{\mathbf{P}}}$
associated with $\left\{ g_{ij}\right\} $ for $\mathcal{\mathbf{P}}$, then  $\gamma
_{ji}^{\prime }=$\ $\Psi \circ \gamma _{ji}$\ are transition functions for  
$\widetilde{\mathcal{\mathbf{P}}}^{\prime }$.

\item  If $\widetilde{\mathcal{\mathbf{P}}}$ and $\widetilde{\mathcal{\mathbf{P}}}^{\prime }$ have transition
functions $\left\{ \gamma _{ij}\right\} $ and $\left\{ \gamma _{ij}^{\prime
}\right\} $ with $\gamma _{ij}^{\prime }=$\ $\Psi \circ \gamma _{ji}$\ then 
$\widetilde{\mathcal{\mathbf{P}}}\simeq _{W}\widetilde{\mathcal{\mathbf{P}}}^{\prime }$.
\end{enumerate}
}}

To show this we observe that:
\begin{enumerate}
\item  Associated with $\{ \gamma_{ij} \}$ is a system $\widetilde{\varphi }_{i}$    of charts 
\begin{eqnarray}
\widetilde{\varphi }_{i} &:&U_{i}\times \Gamma \rightarrow \widetilde{\mathcal{\mathbf{P}}}
\text{, with } \nonumber  \\
\gamma _{ij} &:&U_{i}\smallfrown U_{j}\rightarrow \mathbf{G}\ \nonumber \\
 \text{given
by } 
\widetilde{\varphi }_{i}\widetilde{\varphi }_{j}^{-1}(\widetilde{e}) &=&
\widetilde{e}\cdot \widetilde{\gamma }_{ij} \: .
\end{eqnarray}
Then charts $\widetilde{\varphi }_{i}^{\prime }:$\ $U_{i}\times \Gamma
\rightarrow \widetilde{\mathcal{\mathbf{P}}}^{\prime }$ defined by $\widetilde{\varphi }
_{i}^{\prime } = \Theta \circ \widetilde{\varphi }_{i}$ have transition
functions $\gamma_{ij}^{\prime} : U_{i} \cap U_{j} \rightarrow \mathbf{G}$ given by 
\begin{eqnarray}
\gamma _{ij}^{\prime } &=&\widetilde{\varphi }_{i}\widetilde{\varphi }%
_{j}^{-1}(\widetilde{e}^{\prime }) \nonumber \\
&=&\Theta \circ \widetilde{\varphi }_{i}\circ \widetilde{\varphi }%
_{j}^{-1}\circ \Theta ^{-1}(\widetilde{e}^{\prime }) \nonumber  \\
&=&\Theta \lbrack \widetilde{e}\cdot \gamma _{ij}] \nonumber \\
&=&\widetilde{e}^{\prime }\cdot \Psi (\gamma _{ij})
\end{eqnarray}
\begin{equation}
\Longrightarrow \gamma _{ij}^{\prime }=\Psi (\gamma _{ij}) \: .
\end{equation}
That $\left\{ \gamma _{ij}^{\prime }\right\} $\textit{\ satisfy
consistency relations and }$\rho \circ \gamma _{ij}^{\prime }=g_{ij}$
\ is immediate.

\item  Proof relies here on showing that 
\begin{equation}
\Theta |_{\widetilde{\pi }^{-1}U_{i}}\equiv \widetilde{\varphi }_{i}%
\widetilde{\varphi }_{i}^{-1} \: , 
\end{equation}
 is well defined.
We set, 
\begin{eqnarray}
\widetilde{\varphi }_{1}^{\prime }(u,1) &=&e_{1}^{\prime }, \: \: \:
\widetilde{\varphi }_{1}(u,1)=e_{1} \\
\widetilde{\varphi }_{2}^{\prime }(u,1) &=&e_{2}^{\prime }=e_{1}^{\prime
}\cdot \gamma ^{\prime },\gamma ^{\prime }=\pm 1;\text{ \ }\widetilde{%
\varphi }_{2}(u,1)=e_{2}=e_{1}\cdot \gamma ,\gamma =\pm 1
\end{eqnarray}
\begin{equation}
\widetilde{\varphi }_{1}^{\prime }\widetilde{\varphi }_{2}^{\prime -1}(
\widetilde{e}_{2}^{\prime })=\widetilde{e}_{2}^{\prime }\cdot \gamma
_{12}^{\prime }=\widetilde{e}_{1}^{\prime }\Longrightarrow \gamma
_{12}^{\prime }=\gamma .
\end{equation}
Similarly $\gamma _{_{12}}=\gamma .$ Then on $
\widetilde{\pi }^{-1}\left( U_{i}\smallfrown U_{j}\right) $\textit{\ , }$
\Theta $\textit{\ is well defined if }$\widetilde{\varphi }_{1}\widetilde{%
\varphi }_{1}^{-1}=\widetilde{\varphi }_{2}\widetilde{\varphi }_{2}^{-1}.$%
\begin{eqnarray}
\widetilde{\varphi }_{1}\widetilde{\varphi }_{1}^{-1}(\widetilde{e}_{1}) &=&%
\widetilde{e}_{1}  \nonumber \\
\widetilde{\varphi }_{2}\widetilde{\varphi }_{2}^{-1}(e_{1}) &=&\widetilde{
\varphi }_{2}^{\prime }\widetilde{\varphi }_{2}^{-1}\left( \widetilde{e}
_{2}\cdot \gamma \right) =\widetilde{\varphi }_{2}^{\prime }(u,\gamma )=
\widetilde{e}_{2}^{\prime }\cdot \gamma =\widetilde{e}_{1}^{\prime }\cdot
\gamma ^{\prime }\gamma
\end{eqnarray}
Finally, $\gamma _{12}=\gamma _{12}^{\prime }$ , since $
\Psi $ is an isomorphism : 
\begin{eqnarray}
&\Longrightarrow &\Psi (1)=1,\Psi (-1)=-1 \\
&\Longrightarrow &\gamma ^{\prime }\gamma =\gamma _{12}^{2}=1, 
\end{eqnarray}
which  implies that $\Theta$  is well defined.
\end{enumerate}
Hence we proved our statement.
Now we try to show that if only weak equivalence is imposed on  $\Gamma $%
-structures on $\mathcal{\mathbf{P}}$, then the number of inequivalent $\Gamma $-structures
will no longer be given by $H^{1}(\mathcal{\mathbf{B}},\mathbf{Z}_{2})$ but instead by $H^{1}({
\mathcal{\mathbf{B}}}^{c},\mathbf{Z}_{2})$ .  Note that the covering induced by $p$ is a regular covering.
Hence $p_{\ast }\left( \pi _{1}(\mathcal{\mathbf{B}}^c,\ast )\right) \lhd \pi
_{1}(\mathcal{\mathbf{B}},\ast )$\ and this covering is a  $\mathbf{Z}_{2}$-covering.  We can see
right away
 that the map $\Psi $\ has similar properties on the group action
level.  As an example, take $\mathcal{\mathbf{B}}=RP_{2}$ . Then $\mathcal{\mathbf{B}}^c=\mathbf{S}^{2}.$
\begin{eqnarray}
\pi _{1}(\mathcal{\mathbf{B}}) &=&\mathbf{Z}_{2}, \\
\pi _{1}(\mathcal{\mathbf{B}}^c) &=&1 .
\end{eqnarray}
The space $\mathcal{\mathbf{B}}$ has in this case
two strongly-inequivalent $\mathbf{Pin}^{-}(2)$ structures.  If we
let $w$ denotes the volume element of the Clifford algebra associated to $
\mathbf{Pin}^{-}(3)$, then the two $\mathbf{Pin}^{-}(2)$\ structures are obtained through the
following coverings 
\begin{equation}
\mathbf{Pin}^{-}(3)/\left\{ 1,\pm w\right\} \rightarrow \mathbf{O}(3)/\left\{ 1,-1\right\}
\end{equation}
If we require only weak equivalence, then both structures become equivalent
with 
\begin{eqnarray}
\Psi (w) &=&-w \\
\Psi (1) &=&1 \: ,
\end{eqnarray}
and the above two coverings reduce to a single one, namely 
\begin{equation}
\mathbf{Pin}^{-}(3)\rightarrow \mathbf{O}(3)\rightarrow \mathbf{S}^{2} \: .
\end{equation}
So imposing weak-equivalence is equivalent
to factoring out the effect of
the non-orientability of the manifold $\mathcal{\mathbf{B}}$. Hence we should expect that $H^{1}(
\mathcal{\mathbf{B}}^c,\mathbf{Z}_{2})$\ gives distinct
 physical  $\Gamma $-structures on $\mathcal{\mathbf{B}}$.

\subsubsection*{\textbf{Theorem III}}
{\it{$\mathcal{\mathbf{P}}^{c}$ is a $\Gamma $-structure on $\mathcal{\mathbf{P}}$.  The induced Principal bundle 
$p^{-1}\mathcal{\mathbf{P}}\equiv \mathcal{\mathbf{P}}^{c}$ \ is a trivial double covering for $\mathcal{\mathbf{P}}$. 
Similarly, \ $p^{-1}\widetilde{\mathcal{\mathbf{P}}}\equiv \widetilde{\mathcal{\mathbf{P}}^{c}}$  is a
trivial double covering for $\widetilde{\mathcal{\mathbf{P}}}$\ and $\widetilde{\mathcal{\mathbf{P}}^{c}}$
is a $\Gamma $-structure on $\mathcal{\mathbf{P}}^{c}$ .
}}

\bigskip

We start first by showing that  $\mathcal{\mathbf{P}}^{c}$  is a trivial 
covering of $\mathcal{\mathbf{P}}$. $\mathcal{\mathbf{P}}^{c}$
is by definition the Principal bundle induced by $p$, 
\begin{equation}
\begin{tabular}{llll}
$F:$ & $\mathcal{\mathbf{P}}^{c}$ & $\longrightarrow $ & $\mathcal{\mathbf{P}}$ \\ 
$\overline{\pi }$ & $\downarrow $ &  & $\downarrow \pi $ \\ 
& $\mathcal{\mathbf{B}}^c$ & $\overset{p}{\longrightarrow }$ & $\mathcal{\mathbf{B}}$
\end{tabular}
\end{equation}
Therefore, $\mathcal{\mathbf{P}}^{c}$\ and $\mathcal{\mathbf{P}}$ have the same group 
structure \cite{steenrod}. 
Moreover, we have 
\begin{equation}
\overline{g}_{ji}(x^{\prime })=g_{ji}\left( p(x^{\prime })\right) \text{ \ \
for }x^{\prime }\in \mathcal{\mathbf{B}}^c
\end{equation}
The map F is a 2-1 map.  Now consider the bundle $\mathcal{\mathbf{P}}\times \mathbf{Z}_{2}$  
  defined such that 
\begin{equation}
\begin{tabular}{lll}
$\mathcal{\mathbf{P}}\times \mathbf{Z}_{2}$ & $\overset{\Phi }{\longrightarrow }$ & $\mathcal{\mathbf{P}}^{c}$ \\ 
$\downarrow p^{\prime }$ &  & $\downarrow p$ \\ 
&  &  \\ 
$\mathcal{\mathbf{B}}^c$ & $\overset{id}{\rightarrow }$ & $\mathcal{\mathbf{B}}^c$
\end{tabular}
\end{equation}
 such that 
\begin{eqnarray}
p^{\prime }(u,1) &=&x_{1}^{\prime } \\
p^{\prime }(u,-1) &=&x_{2}^{\prime } \\
p(x_{1}^{\prime }) &=&p(x_{2}^{\prime })=x
\end{eqnarray}
We claim that $\mathcal{\mathbf{P}}\times \mathbf{Z}_{2}\simeq \mathcal{\mathbf{P}}^{c}$\ , i.e., $\Phi $\ is a
bundle isomorphism.  We 
construct $\Phi $  explicitly.  Locally, we have 
\begin{eqnarray}
\Phi &:&\mathcal{\mathbf{P}}\times \mathbf{Z}_{2}\longrightarrow \mathcal{\mathbf{P}}^{c} \nonumber \\
& & \left( x,u(x),1\right) \longrightarrow \left( x_{1}^{^{\prime }},u\right)
 \nonumber \\
& & \left( x,u,-1\right) \longrightarrow \left( x_{2}^{^{\prime }},u\right)
\end{eqnarray}
Therefore, $\overline{\pi }\Phi (x,u,1)=\overline{\pi }($ $x_{1}^{\prime
},u)=x_{1}^{\prime }=p^{\prime }(x,u,1)$\ \ and a similar 
relation holds for $
(x,u,1)$.  Hence, the above diagram commutes and $\Phi $  
carries fibers
to fibers. $\Phi $ can then be considered to be a bundle map 
induced by the
identity.  A similar diagram for 
$\widetilde{\mathcal{\mathbf{P}}}$\ and $\widetilde{
\mathcal{\mathbf{P}}^c}$\ shows 
that $\widetilde{\mathcal{\mathbf{P}}^{c}}$\ is a 
trivial double cover of $
\widetilde{\mathcal{\mathbf{P}}}$\ .

The following diagram gives all possible relations 
among the transition
functions and can be used to prove 
that $\widetilde{\mathcal{\mathbf{P}}^{c}}$ is a  $
\Gamma $-structure on  $\mathcal{\mathbf{P}}^{c}$. 
\begin{equation}
\begin{tabular}{lllll}
$\widetilde{\mathcal{\mathbf{P}}^{c}}$ &  & $\overset{\overline{\Phi }}{\rightarrow }$ & 
& $\mathcal{\mathbf{P}}^{c}$ \\ 
& $\widetilde{\overline{\pi }}\searrow $ &  & $\swarrow $ $\overline{\pi }$
\  &  \\ 
&  & $\mathcal{\mathbf{B}}^c$ &  &  \\ 
$\overline{F}\downarrow $ &  & $\downarrow $ &  & $\downarrow F$ \\ 
&  & $\mathcal{\mathbf{B}}$ &  &  \\ 
& $\nearrow $ &  & $\nwarrow $ &  \\ 
$\mathcal{\mathbf{P}}^{c}$ &  & $\overset{\Phi }{\rightarrow }$ &  & $\mathcal{\mathbf{P}}$%
\end{tabular}
\end{equation}
Considering $\mathcal{\mathbf{P}}^{c}$ 
and $\mathcal{\mathbf{P}} $ as base spaces, the map $\overline{\Phi 
}$\ is induced by $\Phi $.  Hence it is a bundle map.  We 
need to check
that it is equivariant, i.e., 
\begin{equation}
\overline{\Phi }\left( \widetilde{\overline{u}}\cdot \gamma \right) =
\overline{\Phi }(\widetilde{\overline{u}})\cdot \rho (\gamma )\text{ \ \ \ \
\ for all }\gamma \in \Gamma .
\end{equation}
We write $\overline{\Phi }$\ explicitly.  A map that satisfies all the
properties of the above diagram is 
\begin{equation}
\overline{\Phi }(x^{\prime };\widetilde{\overline{u}})=\Phi (x^{\prime };
\widetilde{u}),
\end{equation}
where 
\begin{equation}
p(x^{\prime }) = x
\end{equation}
and $\widetilde{\overline{u}}$\ ,$\widetilde{u}$  are the same pinor
frames.  Since multiplication by $\gamma $ leaves the fiber invariant, it
is trivially true that $\overline{\Phi }$ is an equivariant map since $\Phi $
 is itself equivariant.
\ Next we define the difference class of two structures. $\widetilde{\mathcal{\mathbf{P}}}$ and $\widetilde{\mathcal{\mathbf{P}}}^{\prime }$\ are two $\Gamma $-structures
on $\mathcal{\mathbf{P}}$, where the group actions differ by an isomorphism $\Psi $ .  The {\it
difference class} $\delta (\widetilde{\mathcal{\mathbf{P}}},\widetilde{\mathcal{\mathbf{P}}}^{\prime })$ is
defined to be 
\begin{equation}
\delta _{ji}(x)=\gamma _{ji}(x)\Psi (\gamma _{ji}^{\prime -1}(x)),\text{ \ \
\ \ x}\in U_{ij} \: .
\end{equation}
Similarly, we can define $\overline{\delta }$ for the respective double
covers.\ The difference class 
$\delta (\widetilde{\mathcal{\mathbf{P}}},\widetilde{\mathcal{\mathbf{P}}}^{\prime })$ can be shown to be an element of H$
^{1}(\mathcal{\mathbf{B}},\mathbf{Z}_{2}).$  Similarly, $\overline{\delta }(\widetilde{\mathcal{\mathbf{P}}^{c}},
\widetilde{\mathcal{\mathbf{P}}^{c}}^{\prime })$ is an element of H$^{1}(\mathcal{\mathbf{B}}^c
,\mathbf{Z}_{2}).$ \ 
By definition, we have
\begin{equation}
\delta _{ij}(x)=\gamma _{ij}(x)\Psi (\gamma _{ij}^{\prime }(x)^{-1}) \: ,
\end{equation}
and 
\begin{equation}
\rho (\gamma _{ij})=\rho (\gamma _{ij}^{\prime })=g_{ij} \: .
\end{equation}
This implies that 
\begin{eqnarray}
\rho (\delta _{ij}) &=&1 \\
&\Longrightarrow &\delta _{ij}(x)\in \mathbf{Z}_{2}
\end{eqnarray}
i.e., $\delta _{ij}$\ is in the center of $\Gamma $\ and 
\begin{eqnarray}
(d\delta )(ijk) &=&\delta _{jk}\delta _{ik}^{-1}\delta _{ij}\nonumber \\
&=&\gamma _{ji}\Psi (\gamma _{jk}^{\prime -1})(\gamma _{ik}\Psi (\gamma
_{ik}^{\prime -1}))\gamma _{ij}\Psi (\gamma _{ij}^{\prime -1}) \nonumber \\
&=&\gamma _{jk}\Psi (\gamma _{jk}^{\prime -1})(\Psi (\gamma _{ik}^{\prime
})\gamma _{ik}^{-1})\gamma _{ji}\Psi (\gamma _{ij}^{\prime -1})\nonumber \\
&=&\gamma _{jk}(\gamma _{ij}\Psi (\gamma _{ij}^{\prime -1}))\Psi (\gamma
_{jk}^{\prime -1})(\Psi (\gamma _{ik}^{\prime })\gamma _{ik}^{-1})\text{
since }\delta _{ij}\in C(\Gamma )\nonumber \\
&=&\gamma _{jk}\gamma _{ij}\Psi (\gamma _{ij}^{\prime -1}\gamma
_{jk}^{\prime -1}\gamma _{ik}^{\prime })\gamma _{ik}^{-1} \nonumber \\
&=&1 .
\end{eqnarray}
Hence $\delta _{ij}\in $ $H^{1}(\mathcal{\mathbf{B}},\mathbf{Z}_{2})$\ .  A similar proof works for $
\overline{\delta }_{ij}$. \ The difference class can be used 
to define an equivalence relation among the $\Gamma -$ 
structures. \ In fact, we have 
$\widetilde{\mathcal{\mathbf{P}}}\simeq _{W}\widetilde{\mathcal{\mathbf{P}}}^{\prime }$ iff $\overline{\delta }(%
\widetilde{\mathcal{\mathbf{P}}^{c}},\widetilde{\mathcal{\mathbf{P}}^{c}}^{\prime })=1$ \: . \ To show this, 
suppose that $\widetilde{\mathcal{\mathbf{P}}}$ and $\widetilde{\mathcal{\mathbf{P}}}^{\prime }$\ are weakly
-equivalent, then 
\begin{equation}
\gamma _{ij}(x)=\Psi (\gamma _{ij}^{\prime }(x)) ,
\end{equation}
with \ $\Psi ^{2}=1$.  Moreover we have 
\begin{eqnarray}
\overline{\gamma }_{ij}(x) &=&\gamma _{ij}(p(x^{\prime })) 
\end{eqnarray}
since
 
\begin{equation}
\overline{g}_{ij}( x^{\prime } ) \; =\; g_{ij}(p(x^{\prime })) ,
\end{equation}
which implies that 
\begin{equation}
\overline{\gamma }_{ij}(x^{\prime })=\Psi (\overline{\gamma }_{ij}^{\prime
}(x)) .
\end{equation}
Hence the difference class becomes 
\begin{eqnarray}
\overline{\delta }_{ij}(x) &=&\overline{\gamma }_{ij}(x^{\prime })\Psi (
\overline{\gamma }_{ij}^{\prime -1}(x^{\prime }))\nonumber \\
&=&\overline{\gamma }_{ij}(x^{\prime })(\overline{\gamma }_{ij}^{\prime
-1}(x^{\prime })) \nonumber \\
&=&1 .
\end{eqnarray}
Now suppose that $\overline{\delta }(\widetilde{\mathcal{\mathbf{P}}^{c}},\widetilde{
\mathcal{\mathbf{P}}^{c}}^{\prime })=1$.  Hence there exists $\lambda
_{i} : p^{-1}(U_{i})\rightarrow \Gamma $ such that 
\begin{equation}
d\lambda (ij)=\overline{\delta }(ij) \: .
\end{equation}
This is a \v{C}ech-coboundary condition, therefore we have 
\begin{equation}
\overline{\gamma }_{ij}(x^{\prime })=\lambda _{i}^{-1}(x^{\prime })\Psi (
\overline{\gamma }_{ij}^{\prime }(x^{\prime }))\lambda _{j}(x^{\prime }) \: .
\end{equation}
Now we try to construct locally the bundle isomorphism 
\begin{equation}
\Theta : \widetilde{\mathcal{\mathbf{P}}}\rightarrow \widetilde{\mathcal{\mathbf{P}}}^{\prime }
\end{equation}
such that 
\begin{eqnarray}
\Theta (\widetilde{u}\cdot \gamma ) &=&\Theta (\widetilde{u})\cdot \Psi
(\gamma )
\end{eqnarray}
and 
\begin{eqnarray}
\Phi \circ \Theta &=&\Phi ^{\prime } \; .
\end{eqnarray}
First we have the following commutative diagram: 
\begin{equation}
\begin{tabular}{lllll}
$\widetilde{\mathcal{\mathbf{P}}^{c}}$ &  & $\overset{\overline{\Theta }}{\rightarrow }$
&  & $\widetilde{\mathcal{\mathbf{P}}^{c}}^{\prime }$ \\ 
& $\overline{\Phi }\searrow $ &  & $\swarrow \overline{\Phi }^{\prime }$ & 
\\ 
&  & $\mathcal{\mathbf{P}}^{c}$ &  &  \\ 
$\downarrow $ &  & $\downarrow $ &  & $\downarrow $ \\ 
&  & $\mathcal{\mathbf{P}}$ &  &  \\ 
& $\Phi \nearrow $ &  & $\nwarrow \Phi ^{\prime }$ &  \\ 
$\widetilde{\mathcal{\mathbf{P}}}$ &  & $\overset{\Theta }{\rightarrow }$ &  & $\widetilde{%
\mathcal{\mathbf{P}}^{\prime }}$
\end{tabular}
\end{equation}
It should be clear from this diagram that locally $\Theta $\ and $\overline{\Theta }$\ are
the same.  Hence a construction of $\overline{\Theta }$\ will immediately
give one for $\Theta $ .  Let $V_{i}=\widetilde{\overline{\pi }}
(p^{-1}(U_{i}))$ and define $\overline{\Theta }$  locally by 
\begin{equation}
\overline{\Theta }_{i}:V_{i}
\rightarrow \widetilde{\mathcal{\mathbf{P}}^{c}}^{\prime } 
\end{equation}
\begin{equation}
\widetilde{\overline{u}}\rightarrow \sigma _{i}^{\prime }(\widetilde{
\overline{\pi }}(\widetilde{\overline{u}}))\cdot \lambda _{i}\Psi (\gamma
_{i}(\widetilde{\overline{u}}))
\end{equation}
where $\sigma _{i}^{\prime }$ is a local cross section of $\widetilde{
\mathcal{\mathbf{P}}^{c}}^{\prime }$\ and $\gamma _{i}$\ is an element of $\Gamma $
such that 
\begin{equation}
\widetilde{\overline{u}}=\sigma _{i}(\widetilde{\overline{\pi }}(\widetilde{
\overline{u}}))\cdot \gamma _{i}(\widetilde{\overline{u}})
\end{equation}
If $\gamma \in \Gamma $ , then 
\begin{eqnarray}
\overline{\Theta }_{i}(\widetilde{\overline{u}}\cdot \gamma ) &=&\sigma
_{i}^{\prime }(\widetilde{\overline{\pi }}(\widetilde{\overline{u}}\cdot
\gamma ))\cdot \lambda _{i}\Psi (\gamma _{i}(\widetilde{\overline{u}}\cdot
\gamma )) \nonumber \\
&=&\overline{\Theta }_{i}(\widetilde{\overline{u}}).\Psi (\gamma ) \: .
\end{eqnarray}
This map is well defined globally.  On intersection $V_{i}\smallfrown V_{j}$
\begin{eqnarray}
\sigma _{i}(\widetilde{\overline{\pi }}(\widetilde{\overline{u}}))\cdot
\lambda _{i}\Psi (\gamma _{i}(\widetilde{\overline{u}}) &=&\sigma _{i}(
\widetilde{\overline{\pi }}(\widetilde{\overline{u}}))\gamma _{ji}^{^{\prime
}}\lambda _{i}\Psi (\gamma _{i}) \nonumber \\
&=&\sigma _{i}(\widetilde{\overline{\pi }}(\widetilde{\overline{u}}))\lambda
_{j}\Psi (\gamma _{ji})\Psi (\gamma _{i}) \nonumber \\
&=&\sigma _{i}(\widetilde{\overline{\pi }}(\widetilde{\overline{u}}))\lambda
_{j}\Psi (\gamma _{j}),
\end{eqnarray}
as it should be.

\subsubsection*{\textbf{Theorem IV}}
{\it {Let $\widetilde{\mathcal{\mathbf{P}}}$ be a $\Gamma$-structure on 
$\mathcal{\mathbf{P}}$. $\widetilde{\mathcal{\mathbf{P}}^{c}}$
 is the
corresponding double cover.  Then for each element $\overline{\zeta }\in
H^{1}(\mathcal{\mathbf{B}}^c,\mathbf{Z}_{2})$ there exist a non weakly-equivalent $\Gamma
-structure$ $\widetilde{\mathcal{\mathbf{P}}}^{\prime }.$}}
\bigskip

Let $\overline{\zeta }_{ij}$\ be a representation of 
$\overline{\zeta }$\ and
$\Psi $ is an isomorphism as above.  Define the following functions: 
\begin{equation}
\overline{\gamma }_{ij}(x^{\prime })=\overline{\zeta }_{ij}^{-1}\cdot (
\overline{\gamma }_{ij}^{\prime }(x^{\prime })) \: .
\end{equation}
They clearly satisfy the cocycle condition and hence they form the
transition functions of a principal bundle which we call $(\widetilde{
\mathcal{\mathbf{P}}^{c}}^{\prime },\widetilde{\overline{\pi }}^{\prime },\mathcal{\mathbf{B}}^c
,\Gamma )$ .  Note also that 
\begin{equation}
\overline{\delta }_{ij}=\overline{\gamma }_{ij}(\overline{\gamma }
_{ij}^{^{\prime }})=\overline{\zeta }_{ij} \; ,
\end{equation}
and 
\begin{equation}
\rho (\overline{\gamma }_{ij})=\rho (\overline{\gamma }_{ij}^{\prime })=
\overline{g}_{ij}.
\end{equation}
Now, to get the Principal bundle $(\widetilde{\mathcal{\mathbf{P}}^{c}}^{\prime },
\widetilde{\overline{\pi }}^{\prime },\mathcal{\mathbf{B}}^c,\Gamma )$ we use the
fact that $(\mathcal{\mathbf{B}}^c,p,\mathcal{\mathbf{B}},\mathbf{Z}_{2})$\ is a Principal bundle with transition $
\theta _{ij}(x)\in \mathbf{Z}_{2}$ .  Hence the following diagram commutes: 
\begin{equation}
\begin{tabular}{lll}
$\widetilde{\mathcal{\mathbf{P}}^{c}}^{\prime }$ & $\longrightarrow $ & $\mathcal{\mathbf{B}}^c$
\\ 
&  &  \\ 
& $p^{\prime }\searrow $ & $\downarrow $ \\ 
&  & $\mathcal{\mathbf{B}}$
\end{tabular}
\end{equation}
Therefore $(\widetilde{\mathcal{\mathbf{P}}^{c}}^{\prime },\widetilde{\overline{\pi }}
^{\prime },\mathcal{\mathbf{B}}^c,\Gamma )$\ is a Principal bundle with transition
functions 
\begin{equation}
\psi _{ij}(x)=\theta _{ij}(x)\cdot \overline{\gamma }_{ij}^{\prime
}(x^{\prime }) \: .
\end{equation}
Here, we have used the trivial extension of $\Gamma $\ , i.e., 
\begin{equation}
1\rightarrow \mathbf{Z}_{2}\rightarrow \mathbf{Z}_{2}\otimes \Gamma \rightarrow \Gamma
\rightarrow 1 \: .
\end{equation}
The bundle $ \widetilde{\mathcal{\mathbf{P}}}^{\prime} $  is constructed with 
the same transition functions. 
Therefore we set 
\begin{equation}
\gamma _{ij}^{\prime }(x)=\theta _{ij}(x)\cdot \overline{\gamma }
_{ij}^{\prime }(x^{\prime }) .
\end{equation}

\bigskip

\ This ends our main section which relates the number of inequivalent 
pin structures to the first Cohomology group of the associated 
orientable cover of the underlying non-orientable space.

\section{APPLICATIONS}

\ In this last section, we discuss two examples: the first 
deals with an electron in a nano-circuit. \ The second deals with 
a non-orientable space.

\subsection{Transport in nano-Circuits}

In this section, we follow the notation of Negele and Orland \cite{negele}.
The geometry of the circuit is nontrivial; it has a `hole'. The homotopy group
of the torus is $\pi_{1}\left(  T^{2}\right)  =\mathbf{Z}\oplus \mathbf{Z}$. It
is a two-dimensional surface. However the electrons are not only constrained
to the surface, but they can be also inside. Therefore the geometry of the
circuit is in fact homeomorphic to $D\times \mathbf{S}^{1}$ where $D$ is a disk in
$\mathbb{R}^{2}$, i.e, a simply connected region. Moreover the manifold is
orientable in this case and hence the nontrivial spinor is dictated by the
circle around the hole (see figure 2). \ According to 
our discussion in previous sections, any non-orientability 
will be factored out. \ Hence, the following discussion will 
equally apply to a Mobius band. \ For 
this manifold, there are two
possibilities to define spinors since $H^{1}\left(  \mathbf{M},\mathbf{Z}_{2}\right)
=\mathbf{Z}_{2}$. The vector potential that corresponds to the non-trivial one
differs by an element in the Cohomology class $\lambda$:\cite{petry}%

\begin{equation}
A_{\mu}\rightarrow A_{\mu}-\frac{i\hbar c}{2e}\lambda^{-1}\partial_{\mu
}\lambda ,
\end{equation}
with%
\begin{equation}
\oint\lambda^{-1}\partial_{\mu}\lambda\cdot dx=2\pi i.
\end{equation}
The function $\lambda$ can always be chosen to be defined on the unit circle:%
\begin{equation}
\lambda:M\rightarrow\text{ \ }U(1)\subset\mathbb{C}\nonumber
\end{equation}
Therefore the magnetic flux will change by a discrete value for each closed
path traveled by an electron around the circuit%
\begin{equation}
\oint\left(  A_{\mu}-\frac{i}{2e}\lambda^{-1}\partial_{\mu}\lambda\right)
\cdot dx=\oint A_{\mu}\cdot dx^{\mu}+\frac{hc}{2e}%
\end{equation}
It is interesting to observe that Magnus and Schoenmaker \cite{magnus} had to
postulate the quantization of flux to be able to recover the Landauer-Buttiker
formula for the conductivity. In our case the quantization is automatic for
the non-trivial spin structure. It will be argued below that for this circuit,
it is the configuration with non-trivial spin structure that must be adopted
based on energy arguments. The Frohlich-Studer ($FS$) theory \cite{frohlich} is
a non-relativistic theory that explicitly exhibits the spin degrees of
freedom. This latter theory is $U(1)\times SU(2)$ gauge-invariant. \ The
$SU(2)$ symmetry comes from the spin degrees of freedom of the wave function
of the electron. For a magnetic field $(\mathbf{A}=\frac{1}{2}\mathcal{\mathbf{B}}%
\times\mathbf{r})$ and an electric field in the $z$-direction, the covariant
derivatives in the $FS$ equation take the form:
\begin{equation}
D_{t}=\partial_{t}+e\varphi-ig\mu S_{z}B
\end{equation}
and the spatial derivatives are
\begin{align}
D_{1} &  =\partial_{1}-ieA_{1}+i\left(  -2g\mu+\frac{e\mu}{2m}\right)
ES_{y},\\
D_{2} &  =\partial_{2}-ieA_{2}+i\left(  -2g\mu+\frac{e\mu}{2m}\right)
ES_{x}, \nonumber\\
D_{3} &  =\partial_{3} . \nonumber
\end{align}
In two dimensions with $z=x+iy$, they acquire a simple form
\begin{equation}
D_{+}=D_{1}+iD_{2}=\partial_{z}-ieA_{+}+g^{^{\prime}}ES_{+},
\end{equation}
with%
\begin{equation}
A_{+}=\frac{1}{2}iBz,\qquad S_{+}=S_{x}+iS_{y},\qquad g^{\prime}=-2g\mu
+\frac{e\mu}{2m}.\nonumber
\end{equation}
For a one dimensional ring with radius $a$,  $z=ae^{i\phi}$. Hence, we can
simply set $x^{2}+y^{2}=a^{2}$ without loosing any essential spin-orbit type
terms in the Hamiltonian as it 
is the case in the standard
 formulation
\cite{meijer}. 

Next we comment on a procedure for obtaining the Green's function and the
effective action for a particle interacting with an electromagnetic field
$F_{\mu\nu}=\partial_{\mu}A_{\nu}-\partial_{\nu}A_{\mu}$ based on the proper
time method \cite{schwinger}. \ This is a relativistic method that starts with
the Dirac equation. We just state the results since they exhibit explicit gauge
invariance. For the one-particle Green's function, we have

\begin{align}
G\left(  x^{\prime},x^{\prime\prime}\right)   &  =i\int_{0}^{\infty
}dse^{-im^{2}s}\\
&  \times\left[  -\gamma_{\mu}\left(  x(s)^{^{\prime}}\left|  \Pi_{\mu
}(s)\right|  x(0)^{^{^{\prime\prime}}}\right)  \right.  \nonumber\\
&  +\left.  m\left(  \left.  x(s)^{^{\prime}}\right|  x(0)^{^{^{\prime\prime}%
}}\right)  \right] , \nonumber
\end{align}
where
\begin{align}
\left(  \left.  x(s)^{^{\prime}}\right|  x(0)^{^{^{\prime\prime}}}\right)    &
=-\frac{i}{\left(  4\pi\right)  ^{2}}\exp\left[  ie\int_{x"}^{x^{\prime}%
}dx^{\mu}A_{\mu}(x)\right]  \nonumber\\
& \frac{1}{s^{2}}e^{-L(s)}\exp\left[  i\frac{1}{4}\left(  x^{^{\prime}%
}-x^{^{\prime\prime}}\right)  eF\coth\left(  eFs\right)  \left(  x^{^{\prime}%
}-x^{^{\prime\prime}}\right)  \right]  \nonumber\\
& \times\exp\left[  i\frac{1}{2}e\sigma Fs\right]  ,
\end{align}%
\begin{align}
\left(  x(s)^{^{\prime}}\left|  \pi_{\mu}(s)\right|  x(0)^{^{^{\prime\prime}}%
}\right)   &  =\frac{1}{2}\left[  eF\coth\left(  eFs\right)  -eF\right]
\nonumber\\
&  \times\left(  x^{^{\prime}}-x^{^{\prime\prime}}\right)  \left(  \left.
x(s)^{^{\prime}}\right|  x(0)^{^{^{\prime\prime}}}\right)  ,
\end{align}
and
\begin{equation}
L(s)=\frac{1}{2}tr\log\left[  \left(  eFs\right)  ^{-1}\sinh\left(
eFs\right)  \right]  .
\end{equation}
The trace is over the Dirac matrices.
These expressions are valid in Euclidean space. The phase factor is clearly
isolated in the expression for the Green functions. Hence a non-trivial spin
structure will clearly affect the Green's function of the theory. In
particular the energy will be different in both cases. \ In the following, we
will assume that there is only a magnetic field and no spin-orbit coupling. We
calculate the energy in both cases. In terms of Green's function with one-body
potential, the energy is given by:
\begin{align}
E &  =i\int dx\left(  i\partial_{t}\right)  \left.  G\left(  x,x^{^{\prime}%
}\right)  \right|  _{x=x^{^{\prime}}}\\
&  =\int dx\Psi^{+}\left(  x^{^{\prime}}\right)  i\partial_{t}\left.
\Psi\left(  x\right)  \right|  _{x=x^{^{\prime}}}\nonumber
\end{align}
In Fourier space, we have
\begin{equation}
G_{\alpha\alpha}\left(  \omega,\mathbf{k}\right)  =\frac{\theta\left(
k_{F}-k\right)  }{\omega-\in_{k}+\alpha\mu_{0}H-i\varepsilon},
\end{equation}
where $\alpha=\pm1$, for spin up and spin down. \ For
\begin{figure}
[ptb]
\begin{center}
\includegraphics[
natheight=8.770900in,
natwidth=6.198100in,
height=2.0859in,
width=4.4255in
]
{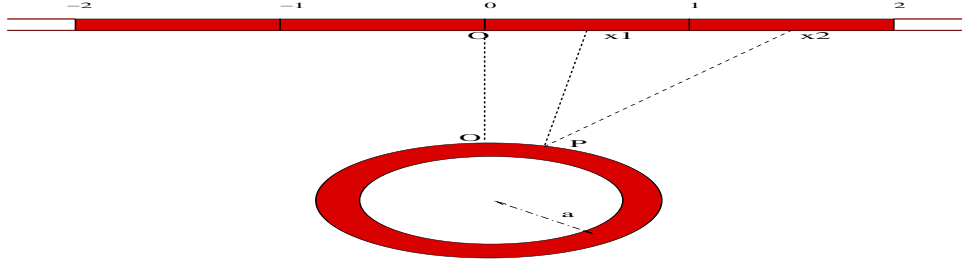}
\caption{Boundary conditions on the Green functions.}
\label{green}
\end{center}
\end{figure}

a periodic lattice with period $d = 2 \pi a$ in the $x$ direction, 
we have \cite{shulman}
\begin{align}
G_{\alpha}\left(  x,x^{^{\prime}}+nd\widehat{x}\right)   &  =-i\theta
\left(  t^{^{\prime}}-t\right)  \int\frac{d^{3}k}{\left(
2\pi\right)  ^{3}}\theta\left(  k_{F}-k\right)  e^{i\mathbf{k\cdot}\left(
\mathbf{x}-\mathbf{x}^{^{\prime}}\right)  }\\
&  e^{-ik_{x}nd}e^{-i\left(  \in_{k}-\alpha\mu_{0}H\right)  \left(
t-t^{^{\prime}}\right)  }. \nonumber
\end{align}
For a regular periodic lattice in Euclidean space, the wave functions are
periodic: this is the configuration that corresponds to the trivial spin
structure. In this case the energy is given by
\begin{align}
E_{0} &  =\sum_{n=-\infty}^{\infty}\sum_{\alpha}\partial_{t}G_{\alpha}\left.
\left(  x,x^{^{\prime}}+nd\widehat{x}\right)  \right|  _{x=x^{\prime}}\\
&  =\frac{1}{2m}\int_{0}^{k_{F}}\frac{dk_{x}}{\left(  2\pi\right)  ^{2}%
}\left(  k_{F}^{4}-k_{x}^{4}\right)  \nonumber\\
&  +\frac{1}{m}\sum_{n=1}^{\infty}\int_{0}^{k_{F}}\frac{dk_{x}}{\left(
2\pi\right)  ^{2}}\cos\left(  ndk_{x}\right)  \left(  k_{F}^{4}-k_{x}%
^{4}\right)  .\nonumber
\end{align}
For a twisted configuration, we have instead the energy:
\begin{align}
E_{t} &  =\sum_{n=-\infty}^{\infty}\left(  -1\right)  ^{n}\sum_{\alpha
}\partial_{t}G_{\alpha}\left.  \left(  x,x^{^{\prime}}+nd\widehat{x}\right)
\right|  _{x=x^{\prime}}\\
&  =\frac{1}{2m}\int_{0}^{k_{F}}\frac{dk_{x}}{\left(  2\pi\right)  ^{2}%
}\left(  k_{F}^{4}-k_{x}^{4}\right)  \nonumber\\
&  +\frac{1}{2m}\sum_{n=\pm2,\pm4,\pm6,..}\int_{0}^{k_{F}}\frac{dk_{x}%
}{\left(  2\pi\right)  ^{2}}\cos\left(  ndk_{x}\right)  \left(  k_{F}%
^{4}-k_{x}^{4}\right)  \nonumber\\
&  -\frac{1}{2m}\sum_{n=\pm1,\pm3,\pm5,..}\int_{0}^{k_{F}}\frac{dk_{x}%
}{\left(  2\pi\right)  ^{2}}\cos\left(  ndk_{x}\right)  \left(  k_{F}%
^{4}-k_{x}^{4}\right)  \nonumber
\end{align}
The difference in energy for typical values of $k_{F}=10^{8}cm^{-1}$ and
$d=100nm$ $(d=10nm)$. In arbitrary units, we have:

\bigskip

$%
\begin{array}
[c]{ccc}%
k_{F} (cm^-1) & d (nm) & E_{t}-E_{0}\\
\\
10^{8} & 100 & -561\\
10^{8} & 10 & -19634
\end{array}
$

\bigskip

Therefore as the size of the ring gets smaller, the nontrivial spin structure
becomes  lower in energy for a critical 
value of the radius. Hence from an energy point of view, the spin 
 will choose to be in the lowest energy state possible that is
compatible with the geometry of the circuit. In this case there will also be a
flux quantization associated with changes in the current. Since we are 
in the ballistic regime, each electron travels in closed paths around the 
circuit. Any change in the number of particles that traveled around the 
torus will give rise to a flux or a vector potential. The current is not 
polarized at zero temperature and hence each pair of electrons with spin 
up and spin up will give a change in flux as that {\it postulated} 
in Ref. \cite{magnus} to recover the Landauer-Buttiker formula 
in non-simply connected circuits with one 
'hole'. Therefore it seems the assumption can 
be proved if the nontrivial spin configuration is taken into account. \ We 
also observe that having twisted leads in the circuit will not change 
the (s)pin structures in this calculation as shown in the 
previous sections.

\begin{figure}
[ptb]
\begin{center}
\includegraphics[
natheight=8.770900in,
natwidth=6.198100in,
height=2.0859in,
width=4.4255in
]
{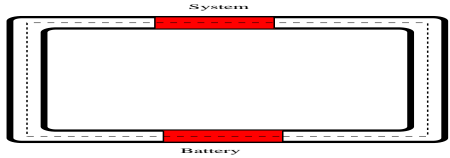}
\caption{A closed nanocircuit: Each electron is assumed to travel around 
the circuit in closed orbits.}
\label{torus}
\end{center}
\end{figure}

\subsection{Spin on a non-orientable space}

In this section we treat non-orientable cases. \ First 
 we take a non-orientable 
manifold, $\mathcal{\mathbf{B}}=\mathbf{S}^{3}/\mathbf{Z}_{2}$, where 
$\mathbf{Z}_{2}=\left\{ (1,T),(1,I)\right\} $. \ From our discussion
above, it was found that
inequivalent ${pin}$  structures were given by $H^{1}(R\times
\mathbf{S}^{3},\mathbf{Z}_{2})$. \ This latter result has been found by different 
methods in \cite{john}.
 \ Finally, we would like to say more about the new adopted
definition for equivalence by going back to the example of the projective
plane that we mentioned earlier.  Here however, we take a more physically
motivated approach.  Let $\psi ^{a},a=1,2$, be a pinor field
 on $\mathbf{RP}_{2}$.
 The structure group of the frame bundle is $\mathbf{O}(2)$.  Let $ \{e_{i}^{a}\}$
be a local frame on the open set $U_{i}.$  The sets $U_{i}$\ cover $\mathbf{RP}_{2}$
 and their intersections are contractible so local sections are always well
defined.  On intersections $U_{i}\smallfrown U_{j}$ we have 
\begin{equation}
e_{i}^{a}=(L_{ij})^{a}e_{i}^{a}\text{, with }L_{ij}\in \mathbf{O}(2) \: .
\end{equation}
On the pinor frame level , we have 
\begin{equation}
\psi _{i}^{a}=(S_{ij})_{b}^{a}\psi _{j}^{a}\text{, with }S_{ij}\in \mathbf{Pin}(2) \; ,
\end{equation}
and $\rho (S)=L$.  We require that
 $\psi ^{\dagger }\psi $ and  $\psi
^{\dagger }\gamma ^{a}\psi $\ transform as a scalar and a 
vector respectively.
The $\gamma ^{a}$ defined here are the Pauli matrices.  From this we get the
following conditions on $S$, 
\begin{eqnarray}
S^{\dagger }S &=&1 \: , \\
S^{\dagger }\gamma ^{a}S &=&L_{b}^{a}\gamma ^{b} \: .
\end{eqnarray}
To find explicit expressions, we need to choose a 
covering. $\mathbf{RP}_{2}$ is
topologically equivalent to a disk with the boundary antipodally 
identified. Next we
 parametrize the boundary with an angle $\theta , 0\preccurlyeq \theta
\prec 2\pi $ . Choose a simple cover for the strip adjacent to the
boundary, we will need at least three open neighborhoods. Non-trivial
transition functions will be needed only as we go along 
the boundary.  They
 are of the form 
\begin{equation}
L=I\cdot e^{2i\theta } \: ,
\end{equation}
where $I$ is a reflection about the first axis.  Using this cover, we find
that $S$ must have the 
form $e^{i\alpha }\gamma ^{1}e^{i\theta \gamma ^{3}}$
 . Imposing boundary conditions on  $\psi (\theta )$ , we find that  
\begin{equation}
e^{i\alpha }=\pm i \: .
\end{equation}
Hence, the two $\mathbf{Pin}(2)$ structures predicted above.  The phase 
factor $
e^{i\alpha }$ is clearly due to the reflection $I$.  Moreover, it is
physically irrelevant and ignoring it amounts to ignoring $I,$ i.e., the
non-orientability of the space. \ Therefore quantum mechanics should 
be studied first on the orientable cover and then projected 
on the configuration  space.

\section{CONCLUSION}

\ In summary, we have given a definition to spin structures
on non-orientable manifolds by going to the 
orientable double cover. \ This allowed us to determine 
the number of inequivalent spin structures using our definition
of equivalence. \ We also showed 
that in the typical 
structure of a nano-circuit, the nontrivial spin configuration 
is probably 
more important than the trivial one at nanometer scale.\ This argument is 
supported indirectly by
the work in ref. \cite{magnus}.\ A convincing proof 
of this statement will 
be to solve the problems with the constraints on the motion of the particle 
explicitly taken into account. \ This is a very difficult problem. We 
believe the
energy argument that we presented is compelling enough to continue looking 
into other aspects which can result from the nontrivial spin 
configuration. \ Smaller non-orientable structures than 
those made by Tanda et al.
should also be possible in the near future and provide an experimental
test of the idea presented here. \ Finally, there is one question that 
we did not discuss in this work and that is related to the nature 
of 'phase transition' at the critical radius of ring between the 
two spin structures. \ This is an interesting question mathematically 
and physically. \ We are not the first to raise this question; 
 Jarosczewicz asked a similar question regarding the spin of 
$\textbf{SU}(2)$ solitons \cite{jaros}. \ To avoid introducing one more 
flavor to quantize the spin, he introduced the idea of a rotating 
soliton which corresponds mathematically to the nontrivial paths 
in $\textbf{SO}(3)$. \ A similar analysis to his may shed some light 
on the physics of our non-trivial spin configurations 
in a ring.

\bigskip

 The author is very grateful to R. Chantrell who made this work possible and 
thanks J. Friedman for initial discussions on this subject.


\begin{thebibliography}{99}


\bibitem{japan} S. Tanda, T. Tsuneta, Y. Okajima, K. Inagaki, K. Yamaya, 
and N. Hatakenaka, Nature \textbf{417}, 397 (2002).

\bibitem{ba} A. P. Balanchandran, G. Marmo, B. S. Skagerstman, 
and A. Stern, Classical Topology and Qunatum States, World 
Scientific (1991).  

\bibitem{thouless}D. J. Thouless, Phys. Rev. B \textbf{27}, 6083 (1983).


\bibitem{rebei} A. Rebei and O. Heinonen, spin currents in the Rashba
 model, submitted to Phys. Rev. textbf{B}.

\bibitem{witten} E. Witten, Nucl. Phys. \textbf{B223}, 422 (1983).


\bibitem{alvarez} L. Alvarez-Gaume and P. Ginsparg, Ann. Phys. (NY) 
\textbf{161}, 423 (1985).


\bibitem{saitoh} E. Saitoh, S. Kasai, and H. Miyajima, J. Appl. 
Phys. \textbf{97}, 10J709 (2005).



\bibitem{dirac}P. A. Dirac, Proc. Roy. Soc. (London) A 117, 610 (1928).






\bibitem{petry} H. B. Petry, J. Math. Phys. \textbf{20}, 231 (1979).

\bibitem{shulman2} L. S. Schulman, Techniques and Applications
of  Path integration. New York, Wiley 1981.

\bibitem{dabrowski}  L. Dabrowski,  Group Actions on Spinors 
 (Bibliopolis, 1988).

\bibitem{cecile} M. Berg, C. DeWitt-Moretti, S. Gwo and E. Kramer, 
Rev. Math. Phys. \textbf{13}, 953 (2001).

\bibitem{atiyah}  M. Atiyah, R.  Bott and A.  Shapiro,   Topology \textbf{3}, Suppl. 1, 3 (1964).

\bibitem{karoubi}  M. Karoubi,  
Ann. Scient. Ec. Norm.\textbf{1}, 161 (1968).

\bibitem{ebner}  D. Ebner,   Gen. Rel. Grav. \textbf{8}, 15
(1977).

\bibitem{eguchi}T. Eguchi, P. B. Gilkey, A. J. Hanson, Phys. Rep. \textbf{66},
213 (1980).


\bibitem{milnor}  J. Milnor,  \ Enseig. Math. \textbf{9}, 198 (1963).


\bibitem{chamblin}  A. Chamblin,  Commun.Math.Phys. \textbf{164}, 65 (1994).

\bibitem{hirzb}  F. Hirzebruch,  Topological Methods in Algebraic Geometry(Springer-Verlag, Berlin, 1966).




\bibitem{spanier}  E. Spanier,   Algebraic Topology  (Springer, New-York,
 1966).




\bibitem{greub}  W. Greub and H. Petry, On the Lifting of Structure 
Groups. \ In:
\ Bleuler,K., Petry,H., Reetz,A. (eds) Differential Geometrical Methods in
Mathematical Physics II.  Proceedings, Bonn 1977, pp. 217-246.  Berlin:
(Springer-Verlag, Berlin,  1978).



\bibitem{steenrod}  N. Steenrod,  The Topology of Fiber Bundles  (Princeton, 1951).

\bibitem{negele} J. Negele and H. Orland, Quantum Many-Partilce Systems, Addison-Wesley, 1988.

\bibitem{magnus} W. Magnus and W. Schoenmaker, J. Math. Phys. 
\textbf{39}, 6715 (1998); Phys. Rev. B \textbf{61}, 10883 (2000).

\bibitem{frohlich} J. Frohlich and U. M. Studer, Rev. Mod. Phys. 
\textbf{65}, 733 (1993).



\bibitem{meijer} F. E. Meijer, A. F. Morpurgo and T. M. Klapwijk, Phys. Rev. 
\textbf{B} 66, 033107 (2002).


\bibitem{schwinger} J. Schwinger, Phys. Rev. \textbf{82}, 664 (1951).
 
\bibitem{shulman} L. S. Schulman, Phys. Rev. \textbf{188}, 1139 (1969).

\bibitem{john}  J. Friedman,  Class.Quantum.Grav.\textbf{12}, 2231
(1995).


\bibitem{jaros}T. Jaroszewicz, Phys. Rev. D \textbf{44}, 1311 (1991).


\end{thebibliography}
\end{document}